\documentclass{aa}  %

\usepackage{graphicx}
\usepackage{txfonts}
\usepackage{longtable}
\begin{document}

   \title{The intense production of silicates during the final AGB phases of intermediate mass stars}

%  \subtitle{I. Gas }

   \author{E. Marini\inst{1}, F. Dell'Agli\inst{1}, D. Kamath\inst{2,3,1}, P. Ventura\inst{1}, L. Mattsson, T. Marchetti\inst{4}, D. A.  Garc{\'\i}a-Hern{\'a}ndez\inst{5,6}, \\ R. Carini\inst{1}, M. Fabrizio\inst{1,7} and S. Tosi\inst{1,8}
          }

   \institute{INAF, Observatory of Rome, Via Frascati 33, 00077 Monte Porzio Catone (RM), Italy \and 
   School of Mathematical and Physical Sciences, Macquarie University, Sydney, NSW, Australia \and
   Astronomy, Astrophysics and Astrophotonics Research Centre, Macquarie University, Sydney, NSW, Australia \and 
   European Southern Observatory, Karl-Schwarzschild-Strasse 2, D-85748 Garching bei M\"unchen, Germany \and 
   Instituto de Astrof\'isica de Canarias (IAC), E-38200 La Laguna, Tenerife, Spain \and
  Departamento de Astrof\'isica, Universidad de La Laguna (ULL), E-38206 La Laguna, Tenerife, Spain \and
  Space Science Data Center, via del Politecnico snc, 00133 Roma, Italy \and
  Dipartimento di Matematica e Fisica, Universit\`a degli Studi Roma Tre, Via della Vasca Navale 84, 00100, Roma, Italy
   }

   \date{}

% \abstract{}{}{}{}{} 
% 5 {} token are mandatory

 \abstract
   {The formation of silicates in circumstellar envelopes of stars
    evolving through the asymptotic giant branch (AGB) is still highly debated
    given the uncertainties affecting stellar evolution modelling, the description
    of the dust formation process, and the capability of silicate grains to
    accelerate stellar outflows via radiation pressure.}
   {We study the formation of dust in the winds of intermediate mass
   (${\rm M} \geq 4~{\rm M}_{\odot}$) stars of solar metallicity while evolving through
   the AGB phase. We tested the different treatments of the mass-loss mechanism by this class of stars, with the aim of assessing their contribution to the general enrichment of silicates of the interstellar medium of galaxies and, on more general grounds, to the silicates budget of the Universe.}
   {We consider a sub-sample of AGB stars, whose spectral energy distribution (SED) is characterised by deep absorption features at $10~\mu$m and $18~\mu$m, which can be regarded as the class of stars providing the most relevant contribution to the silicates' production across the Universe. Results from stellar evolution and dust formation modelling were used to fit the observed SED and to reproduce, at the same time, the detected pulsation periods and the derived surface chemical composition. This analysis leads to the derivation of tight constraints on the silicates' production rates experienced by these sources during the final AGB stages.}
   %We study the spectral energy distributions of a sample of sources    characterized by deep absorption features associated to silicates,    which can be considered as representative of stars providing the    most relevant contribution to the silicates enrichment of the   interstellar medium. The characterization of the stars considered    is based on combination of analysis of the spectral energy    distribution and on information regarding pulsation period    and the surface chemical composition, which allows comparison    with results from stellar evolution and dust formation modelling.}
   {Two out of the four sources investigated are interpreted as stars
   currently undergoing hot bottom burning (HBB), evolving through 
   phases close to the stage when the mass-loss rate is largest. The
   remaining two stars are likely evolving through the very final 
   AGB phases, after HBB was turned off by the gradual consumption of
   the convective mantle. Mass-loss rates of the order of 
   $1\times 10^{-4}~{\rm M}_{\odot}/$yr to $2\times 10^{-4}~{\rm M}_{\odot}/$yr are required when looking
   for consistency with the observational evidence. These results
   indicate the need for a revision of the silicate yields by intermediate
   mass stars, which are found to be $\sim 3$ times higher than previously
   determined.}
   {}


   \titlerunning{The intense production of silicates from AGBs}
   \authorrunning{E. Marini et al.}
   \maketitle
%
%-------------------------------------------------------------------

\section{Introduction}
Silicate dust grains have been detected in a wide variety of environments, ranging from nearby 
protoplanetary disks \citep{maaskant15} to active galactic nuclei \citep[AGNs,][]{xie17} and distant quasars \citep{pennock22}. These particles play an important role in the cosmic life cycle of matter \citep{henning10} since they regulate the thermal structure of the dense and cold phases of the interstellar and
circumstellar dust populations. Furthermore, silicate grains contribute to the
interstellar extinction and emit thermal radiation at IR and millimetre wavelengths. Their mid-IR
spectral features have an important diagnostic value for constraining both the chemical composition
of dust and the grain size distribution. The analysis of these features provides information about the
thermal and density structure of circumstellar disks and envelopes and the toroidal structures
around AGNs \citep{granato94, shi06, xie17}.

Asymptotic giant branch (AGB) stars are probably the most efficient manufacturers of silicates in
the Universe \citep{fg06}. Those providing the most relevant contribution to the overall silicate
budget of the interstellar medium are those with a mass in the $4~{\rm M}_{\odot} \leq {\rm M} \leq 8~{\rm M}_{\odot}$
range, known as intermediate mass stars \citep{fg06, ventura14}. This class of objects has attracted a great deal of interest from the scientific community since it was shown that their contribution to dust production in the Universe cannot be neglected, even in early epochs \citep{rosa09, rosa17}. 

To be able to study dust production by AGB stars, some research groups presented updated models of the AGB phase in which the evolution of the central star is coupled to the description of the dust formation process and the relative impact on the wind \citep{ventura12, ventura14, nanni13, nanni14}, 
following the schematisation proposed by the Heidelberg group \citep{fg02, fg06}. These models have been successfully applied to study the evolved stellar populations of the Magellanic Clouds \citep{flavia14b, flavia15a, flavia15b, nanni16, nanni18, nanni19b} and other galaxies in the Local Group \citep{flavia16, flavia18a, flavia19}.

Despite these steps forward, the contribution from intermediate mass stars to the overall silicate budget of individual galaxies, and more generally of the Universe, is still highly debated. While we expect negligible production in metal-poor environments, the amount of silicates produced by stellar populations of sub-solar, solar, and super-solar chemical composition  is still extremely uncertain, and the results found in the literature differ considerably \citep{schneider14, ventura14, ventura20}. This is partly due to the uncertainties in AGB modelling, which concern the luminosities and the mass-loss rates reached by intermediate mass AGBs \citep{karakas14}, and the possibility that these stars reach the C-star stage towards the end of the AGB lifetime \citep{ventura18}. A further source of uncertainty, owing to the formation process of dust particles, is 
whether chemisputtering or vaporisation is the mechanism responsible for the destruction of silicate grains
\citep{nanni13}. Finally, while the formation of carbon dust was shown to lead to efficient acceleration of the outflow via radiation pressure on solid grains, the formation of silicate particles results in insufficient radiative pressure to drive a wind \citep{hoefner08}; this introduces further uncertainties on the dynamical description of the 
outflow of oxygen-rich AGBs, which has a bearing on the silicate formation process. We note that very recent results for M-type AGB stars \citep{christer23} obtained with the new T-800 code \citep[see][for more details]{sandin20} suggest that even accounting for gas-dust drift to the picture would not lead to higher mass-loss rates for oxygen-rich stars.

%{\bf The systematic study by \citet{bladh19} has also shown that M-type AGB stars rarely form outflows of more than $\dot{M} \sim 10^{-5}\,M_\sun$~yr$^{-1}$. In addition, very recent results for M-type AGB stars (Sandin \& Mattsson, priv. comm.) obtained with the new T-800 code \citep[see][for more details]{sandin20}, suggest that adding gas-dust drift to the picture leads to generally orders of magnitude lower mass-loss rates than those found by \citet{bladh19}.}

In this work we focus on a sample of Galactic AGB stars, whose spectral
energy distributions (SEDs) exhibit deep absorption features at $10~\mu$m and $18~\mu$m, which witness the
presence of large amounts of silicates in the circumstellar envelope. %These spectra, processed and renormalized by \citet{sloan03}, are included in the atlas of spectra from the SWS on \textit{ISO}, covering the 2.4-45.4 $\mu m$ wavelength range. We consider 4 of these objects, for which the pulsation periods and the surface chemical composition, in particular the $^{12}$C$/^{13}$C ratio, are available; this information is crucial characterise these sources, in terms of mass and formation epoch of the progenitors, as well as the mineralogy and radial distribution of the dust in their circumstellar envelope.
These stars descend from intermediate mass progenitors, and can therefore be considered as representative of the stars providing the largest contribution to the overall release of silicates in the interstellar medium. A thorough characterisation of these objects proves crucial to 
deduce the rates at which mass is lost from intermediate mass AGBs, the efficiency of the dust
formation, and -- on more general grounds -- to assess
the silicate budget expected in galaxies and in the Universe.

To this aim, we followed the approach by \citet{ester20, ester21} to study dusty, evolved stars in
the Large Magellanic Cloud (LMC), where results from stellar evolution and dust formation modelling were used to
build a sequence of synthetic spectra, which allow us to describe how the SED of stars with different 
mass evolves during the AGB phase. Comparison with the observations was used in an
attempt to identify the evolutionary phase that is currently experienced by the individual sources, to
characterise the progenitors, and to validate the model adopted to describe
the dust formation process in the winds of intermediate mass AGB stars.

The present investigation, based on the hypothesis that the four objects examined here 
are single stars, offers a complementary characterisation to the one suggested by \citet{decin19},
who propose that two out of the four stars investigated belong to binary systems, on the basis
of Atacama Large Millimeter/submillimeter Array (ALMA) observations of low-excitation rotational lines of $^{12}$CO. This issue is connected
to the more general argument of understanding the origin and the current evolutionary
status of extreme OH$/$IR stars in the Galaxy.

The paper is structured as follows. The sample of stars discussed in the present work is
described in section \ref{sample}. In section \ref{models} the numerical and physical
ingredients used to model the evolution of the star and to produce the synthetic SED
are given. The conclusions drawn from the analysis of the SED of the sub-sample examined
are given in section \ref{sed_fitting}. The main aspects of the AGB evolution of
intermediate mass stars is given in section \ref{agbev}, while the role of the
description of mass loss is discussed in section \ref{mloss}. Section \ref{sources} is devoted to
the characterisation of the sources investigated here. The implications of
the present study for the dust yields of intermediate mass stars is commented on in section
\ref{yields}. Finally, the conclusions are given in section \ref{conc}.

\section{The selected sample}
\label{sample}

The present work is based on the sample of spectra reprocessed and re-normalised by \citet{sloan03}, which are included in the archive for the Infrared Space Observatory (\textit{ISO}). This database contains the observations of a wide variety of sources taken with the Short Wavelength Spectrometer (SWS) in full-scan mode, covering the 2.4-45.4 $\mu m$ wavelength range.
The stars observed have been classified in different classes according to the morphology of their spectra by \citet{kraemer02}. For the present study, we focus on those classified as AGB stars and exhibiting the deepest silicate absorption features at $10~\mu$m and $18~\mu$m and we consider four of these objects for which the pulsation periods and the surface chemical composition, in particular the $^{12}$C$/^{13}$C ratio \citep{kate13}, are available.\ This information is crucial to characterise these sources in terms of the initial mass and formation epoch of the progenitors, as well as the mineralogy and radial distribution of the dust in their circumstellar envelope.
The abundances' and pulsation periods' measurements are even more essential in this context since these stars are almost invisible in the optical; this strongly affects the possibility of an accurate measurement of their parallaxes by \textit{Gaia}.

\section{Numerical and physical inputs}
\label{models}
The study presented here is based on results from stellar evolution modelling (see section~\ref{models_evolution}), which was
required to calculate the time variation of the main physical and chemical quantities of
intermediate mass stars during the AGB lifetime. Dust formation modelling (see section~\ref{models_dust}) was used to
calculate the dust production rate of the various dust species and the optical depth of
the star during the AGB evolution. Finally, results from radiation transfer were used to
build the synthetic SED, which is to be compared with the observations. In the following sub-sections, we give
a brief description of the numerical and physical ingredients adopted for each of the
three tasks above.

\subsection{Stellar evolution modelling}
\label{models_evolution}
We calculated evolutionary sequences of ${\rm M} \geq 4~{\rm M}_{\odot}$ stars of solar metallicity,
evolved from the pre-main sequence until the almost complete ejection of the external mantle.
To this aim we used the ATON code for stellar evolution, described in detail in \citet{ventura98}.
We briefly discuss the physical ingredients most relevant to this work, namely the description
of convection and of mass loss.

The temperature gradient within regions unstable to convection was calculated via the Full Spectrum 
of Turbulence (FST) model \citep{cm91}. Nuclear burning and mixing of chemicals are self-consistently coupled by means of a diffusive approach, according to the schematisation by \citet{clou76}.
The overshoot of convective eddies within radiatively stable regions was modelled by assuming that the velocity of convective elements decays exponentially beyond the neutrality point, which is fixed via the Schwartzschild criterion. The e-folding distance of the velocity decays during the hydrogen and helium core burning 
phases and during the AGB phase was taken as 0.02${\rm H}_{\rm P}$ and 0.002${\rm H}_{\rm P}$, respectively 
(${\rm H}_{\rm P}$ is the pressure scale height at the formal convective border). The latter values reflect the calibrations based on the observed width of the main sequence of open clusters and on the luminosity function of the LMC carbon stars, discussed in \citet{ventura98} and \citet{ventura14}, respectively.

Mass loss during the evolutionary stages preceding the AGB phase was modelled by the classic
Reimers' formula. The treatment of pre-AGB mass loss is of minor importance here as little mass is
expected to be lost during the red giant branch phase.
Regarding the AGB evolution, to model mass loss, we considered the treatments by \citet{blocker95} (hereafter Blo95) and the classic description by Vassiliadis \& Wood (1993, hereafter VW93). In the first case, the mass-loss rate was modelled with the formula
$$
\dot{\rm M}=4.83\times 10^{-22}\eta_R{\rm L}^{3.7}{\rm R}{\rm M}^{-3.1}
\eqno.(1)
$$

The free parameter entering equation (1) was set to $\eta_R=0.02$, following the
calibration based on the luminosity function of lithium-rich stars in the LMC, given in \citet{ventura00}.
The VW93 mass-loss prescription is simply eq. 5 in VW93. The period of the 
star, entering the VW93 recipe, was calculated by means of equation 4 in VW93. During the super-wind
phase, we assumed $\dot{\rm M}=\beta ({\rm L}/{\rm c}{\rm v}_{\rm exp})$, following
\citet{vw93}. We note that $\beta$ represents the average number of scattering of a single photon released by the stellar photosphere by dust particles, under the assumption that the wind is driven by the radiation incident on dust.
A natural first assumption is $\beta=1$, which corresponds to a single scattering. On the
other hand, $\beta$ was shown to increase with opacity \citep{lefevre89}, which is consistent with the
possibility that photons are exposed to multiple scattering when travelling through a high
dust density medium. Detailed radiative transfer computations by \citet{lefevre89} show that
$\beta \sim 2$ when $\dot{\rm M} \sim 10^{-4}~\dot{\rm M}_{\odot}/$yr. Observational 
evidences that the largest mass-loss rates experienced by intermediate mass AGBs exceed
the single scattering radiation limit are presented in \citet{vanloon99}.
We explore the $1\leq \beta \leq 3$ range consistent with the study by \citet{knapp86}.

\subsection{Dust production}
\label{models_dust}

The analysis of the IR spectra of the extremely obscured oxygen-rich stars presented in this work requires knowledge of the dust that formed in the circumstellar envelope. To this aim, we modelled the dust formation and the growth of the dust particles  following \citet{fg06}, according to which the dust forms in a stationary wind that is expanding radially from the photosphere of the star. We refer readers to \citet{ventura12} for all the relevant equations. 
The input parameters required to describe dust formation at a given evolutionary phase of the star are mass, luminosity, mass-loss rate, effective temperature, and surface chemical composition. All of these quantities were obtained by the modelling of the central star, as described earlier in this section.

The key factor affecting the mineralogy of the dust that formed is the C$/$O ratio, thanks to the high stability of the CO molecule \citep{sharp90}. For the winds of M-type AGB stars, we assume that the dust species that formed are silicates, alumina dust (Al$_2$O$_3$), and solid iron.
While Al$_2$O$_3$ is the most stable compound, forming closer ($\sim 2$ stellar radii) to the surface of the star \citep{flavia12}, the species that formed in the largest quantities, thus providing the dominant contribution to the acceleration of the outflow via radiation pressure, are silicates.

The dynamics of the wind is described by the momentum equation, where the acceleration is determined by the competition between gravity and radiation pressure on the newly formed dust grains. 
To calculate the extinction coefficients entering the momentum
equation and the synthetic spectra, as described
in next section, we used the following optical constants:
%The coupling between grain growth and wind dynamics is given by the extinction coefficients, which describe absorption and scattering of the radiation by dust particles. 
for Al$_2$O$_3$ and solid iron, we used the extinction coefficients by \citet{koike95} and \citet{ordal88}, respectively; for amorphous silicates, we considered \citet{draine84}, \citet{suh99}, \citet{oss92}, and \citet{dor95}; whereas, for crystalline silicates, we adopted both the coefficients by \citet{jager94} and the crystalline olivine provided by the DUSTY library, which in turn was taken from the Jena-St. Petersburg database of optical constants\footnote{https://www.astro.uni-jena.de/Laboratory/OCDB/}.

The amount of dust that formed is tightly connected with the mass-loss rate \citep{fg06}. This is because according to the mass conservation law, the mass-loss rate is proportional to the density
of the wind \citep[see equation 4 in][]{ventura12}: 
the larger the $\dot{\rm M}$, the larger the gas density of the outflow will be and also the higher the number of 
molecules available to condense into dust will be. Therefore, for a given surface chemical composition, 
the evolutionary phases during which the stars experience the highest mass-loss rates are when they produce the largest quantities of dust.
%produce the largest quantities of dust are when they experience the highest mass-loss rates.
The result of the modelling of dust formation in the outflow is  the thermodynamical and chemical stratification of the wind, the dust composition, the sizes of the different dust species, the asymptotic velocity, and the optical depth, which we calculated at the wavelength $\lambda = 10~\mu$m ($\tau_{10}$).

\subsection{Spectral energy distribution}
\label{models_sed}
The characterisation of the stars presented in this work requires 
the interpretation of their observed SED, considering that the depth and the shape of the different 
spectral features is extremely sensitive to the mineralogy and to the 
amount of dust present in the circumstellar envelope. 
This task is based on the comparison of the observed spectrum
with the synthetic SED, calculated by means of the code DUSTY \citep{nenkova99}. 

DUSTY calculates the SED of the radiation released from a stellar source, after being
scattered, absorbed, and re-emitted by a dusty region, spherically distributed around the central star. 
The input necessary for the calculation of the synthetic SED are the 
mineralogy and grain size distribution, the optical depth $\tau_{10}$, 
the dust temperature at the inner boundary of the dusty region, and the radial 
distribution of the gas density. All of these inputs are provided on the basis of the results obtained by dust formation modelling as described in the previous sub-section. The spectrum emerging from the photosphere of 
the star\footnote{i.e. the SED found by interpolation in surface
gravity, effective temperature, and C/O ratios among NextGen atmospheres \citep{NextGen} of solar metallicity.} must also be indicated; however, in the present case, the results are 
substantially unaffected by this latter input. This is because the dusty region is 
optically thick, thus the reprocessing of the radiation coming from the 
central star keeps no memory of the incoming radiation from the
stellar photosphere. This allows to use DUSTY in the modality in which the mass loss is taken from the results of stellar evolution modelling and so the density distribution of the outflow. From this we obtained insights about the dust mineralogy and the optical depth of the 
individuals sources, by looking for consistency between the observed
and synthetic SED.

To put further constrains on the physical properties of the central stars,
for each source we also ran DUSTY in the modality where the density
distribution is not provided a priori, rather it is derived by the code by means of the hydrodynamic calculations applied to the wind. This allows for a self-consistent determination of the mass-loss rate and of the terminal outflow velocities. A more exhaustive description is found in \citet{nenkova99}.

\begin{figure*}
\begin{minipage}{0.48\textwidth}
\resizebox{1.\hsize}{!}{\includegraphics{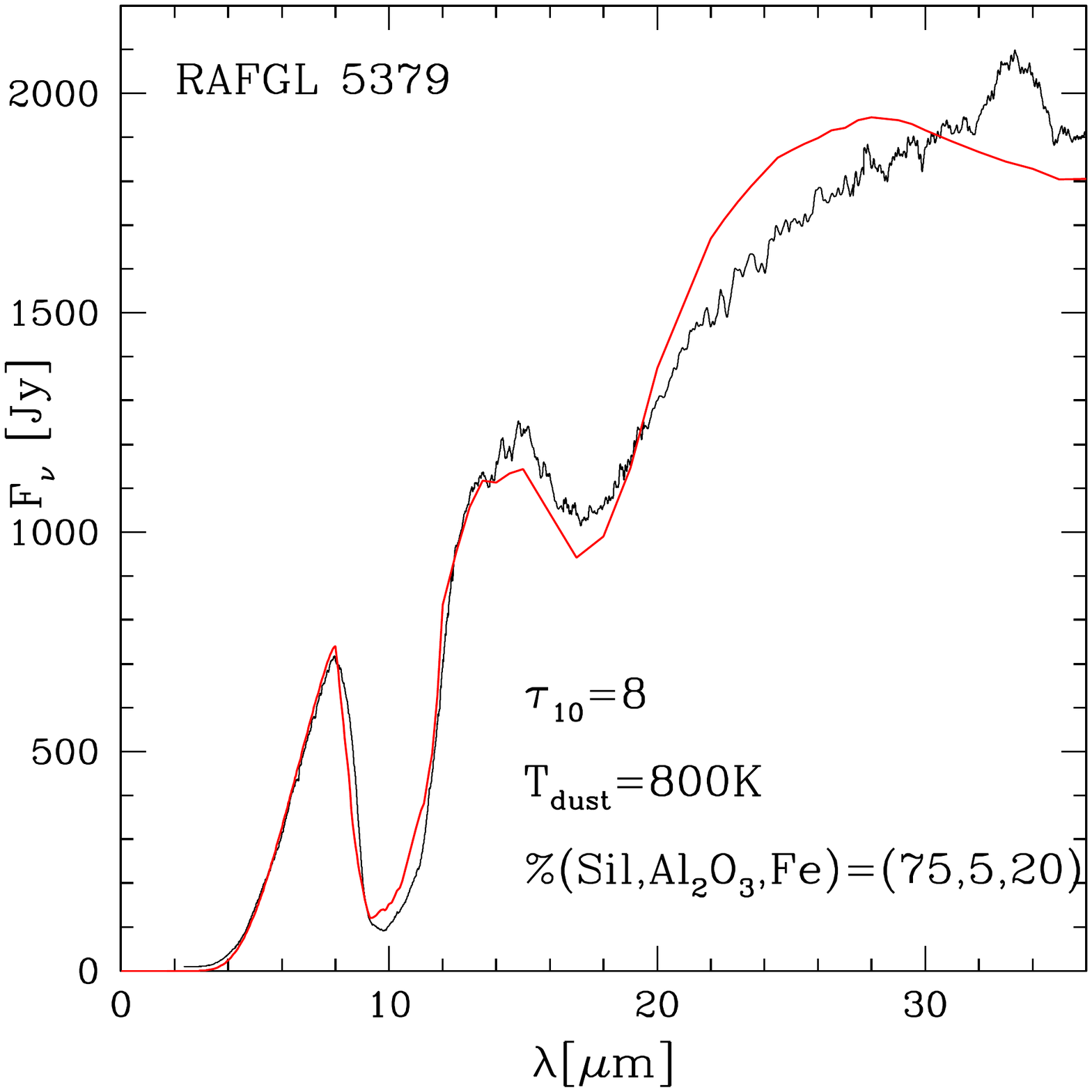}}
\end{minipage}
\begin{minipage}{0.48\textwidth}
\resizebox{1.\hsize}{!}{\includegraphics{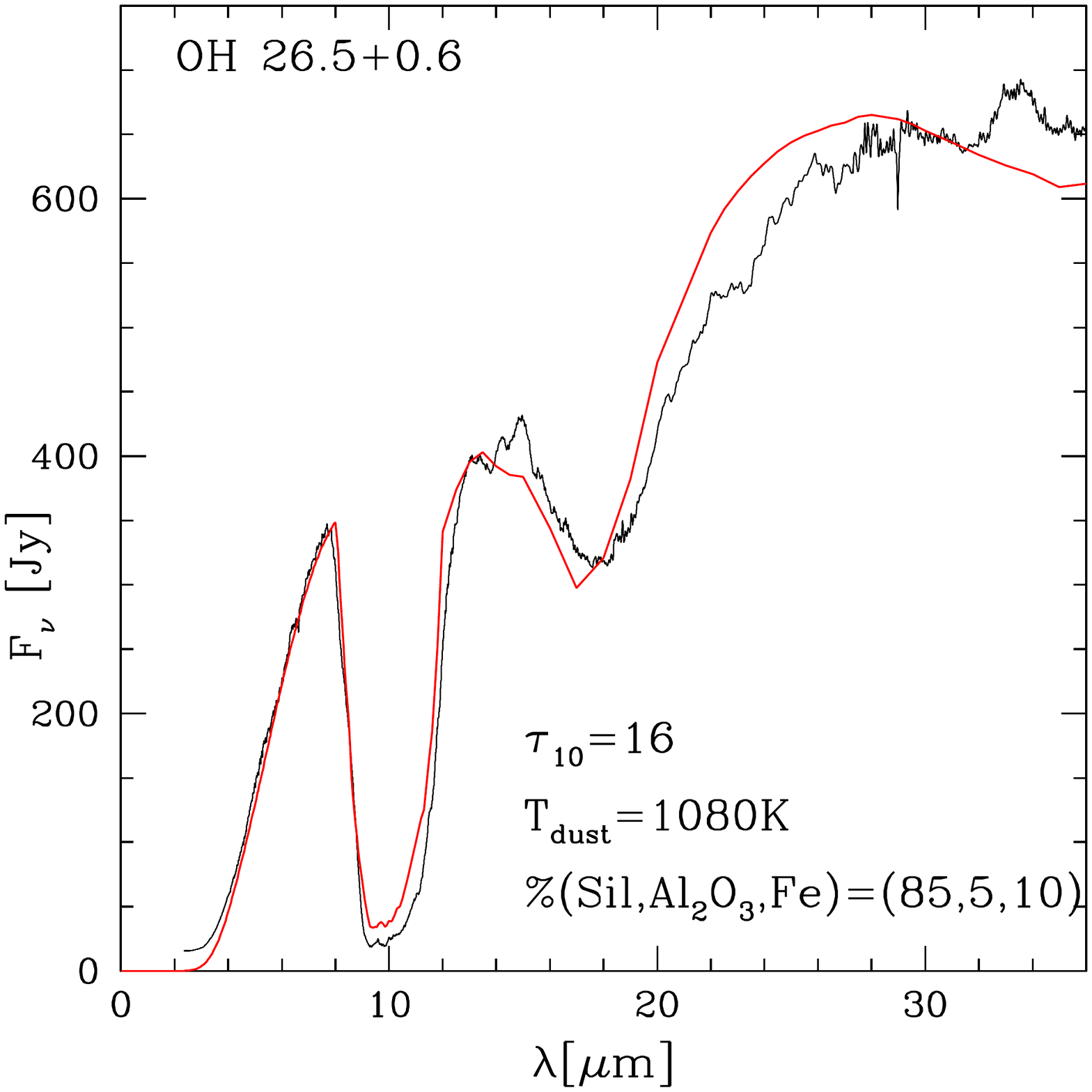}}
\end{minipage}
\vskip-80pt
\begin{minipage}{0.48\textwidth}
\resizebox{1.\hsize}{!}{\includegraphics{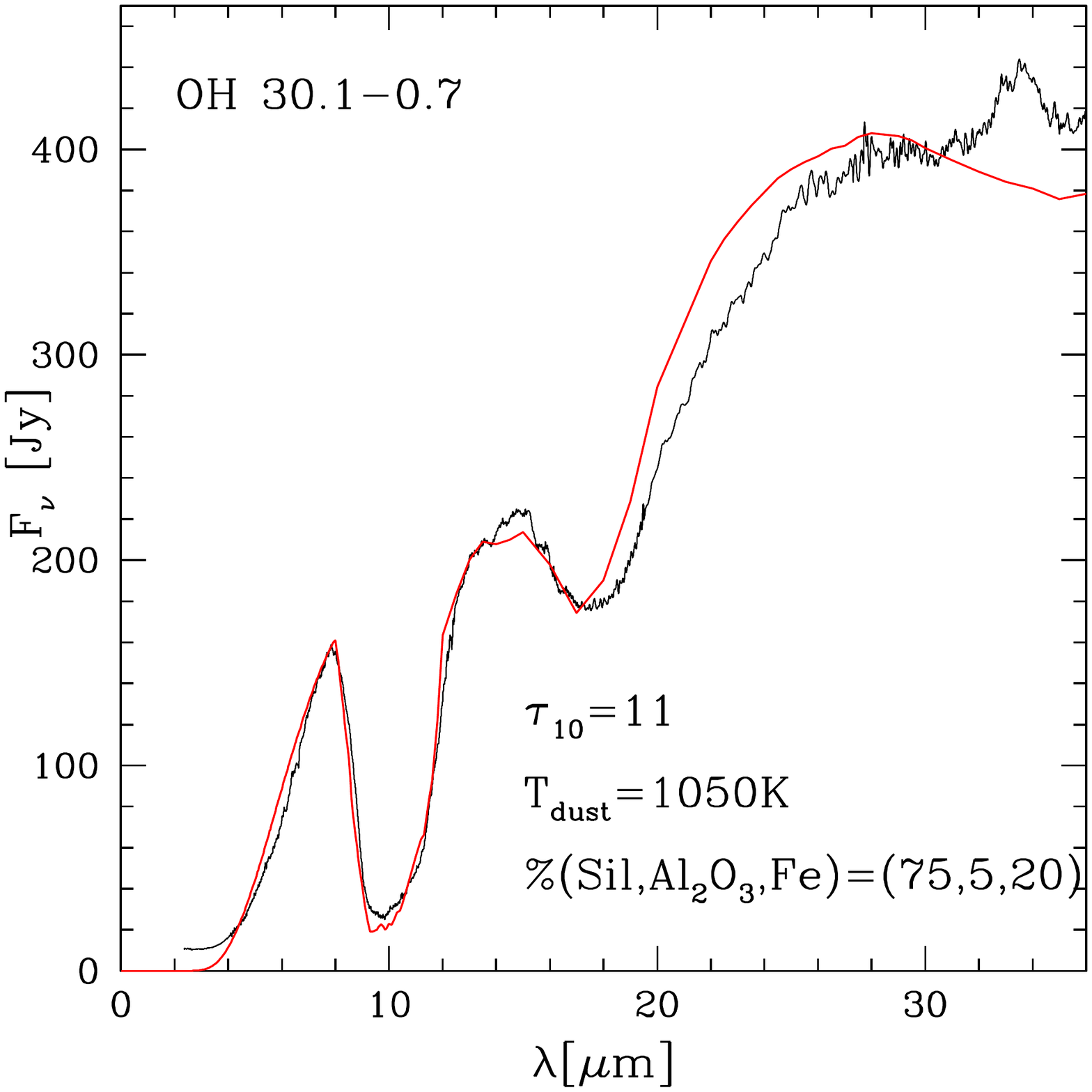}}
\end{minipage}
\begin{minipage}{0.48\textwidth}
\resizebox{1.\hsize}{!}{\includegraphics{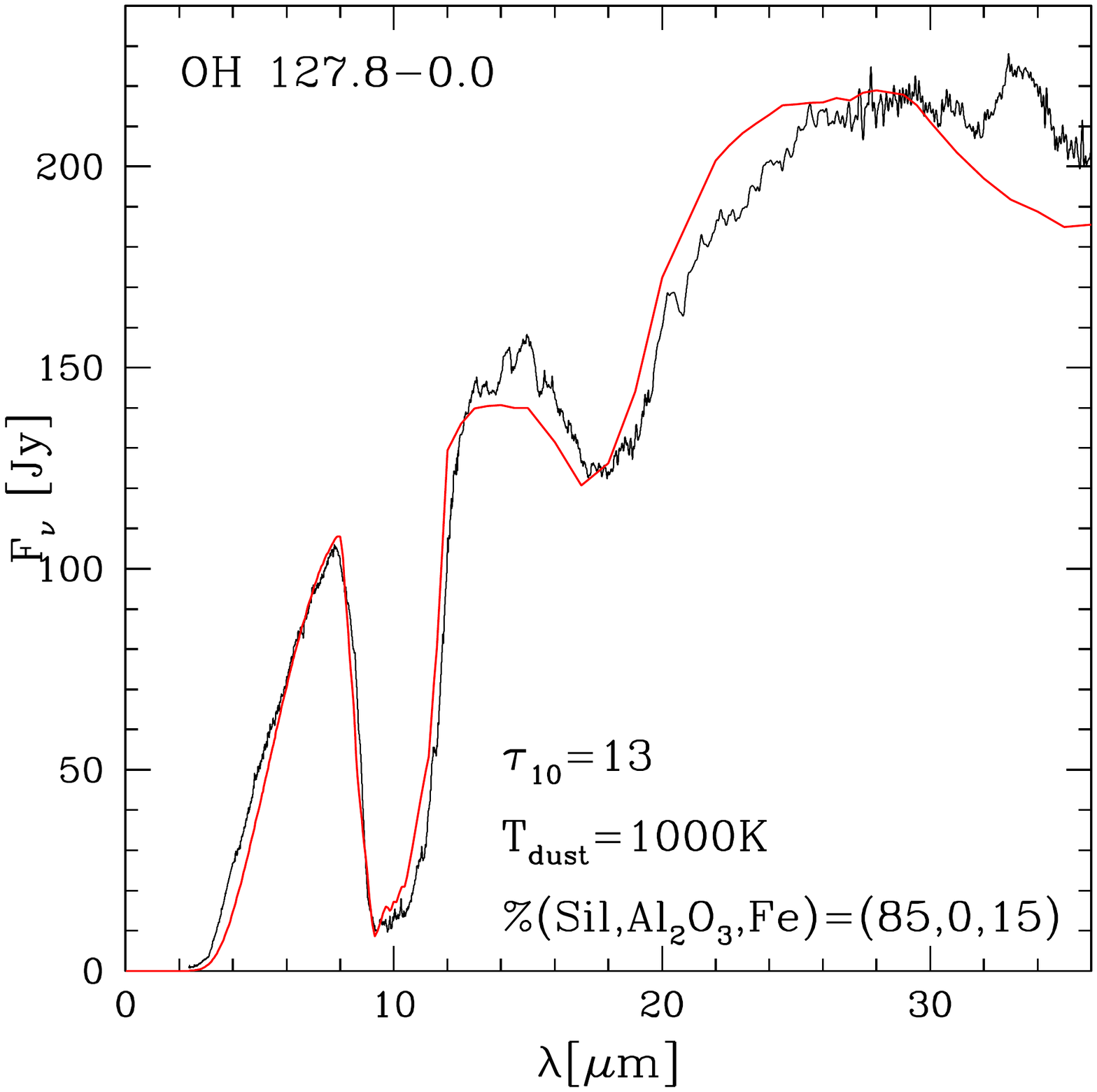}}
\end{minipage}
\vskip-50pt
\caption{ \textit{ISO} spectra (black lines) of the four stars considered in this work. The best fit, indicated with
the red line, was obtained by assuming $\tau_{10}$, T$_{dust}$, and dust composition indicated in the panels.}
\label{fsed}
\end{figure*}

\section{The analysis of the \textit{ISO} spectra}
\label{sed_fitting}

Fig.~\ref{fsed} shows the interpretation of the \textit{ISO} spectra of the four stars analysed in this work.
As described in section~\ref{models_sed}, the identification of the synthetic SED that best reproduces the \textit{ISO} spectrum leads to a robust derivation of the wind properties (e.g. optical depth and mass-loss rates), which we used to characterise the sources.
For the stars considered in this work, we find values of optical depth 8<$\tau_{10}$<16, which is required to reproduce the depth of both the silicate features at 10 and 18$~\mu$m, the slope of the continuum in the spectral region $\lambda$<8$~\mu$m, and the large IR emission at $\lambda$>12$~\mu$m. This part of the spectrum, as well as the depth of the 18$~\mu$m feature also allow for the determination of the dust temperature $T_{dust}$, which is found to be in the range 800-1100 K, as reported in Fig.~\ref{fsed}. 

\begin{figure}
\begin{minipage}{0.48\textwidth}
\resizebox{1.\hsize}{!}{\includegraphics{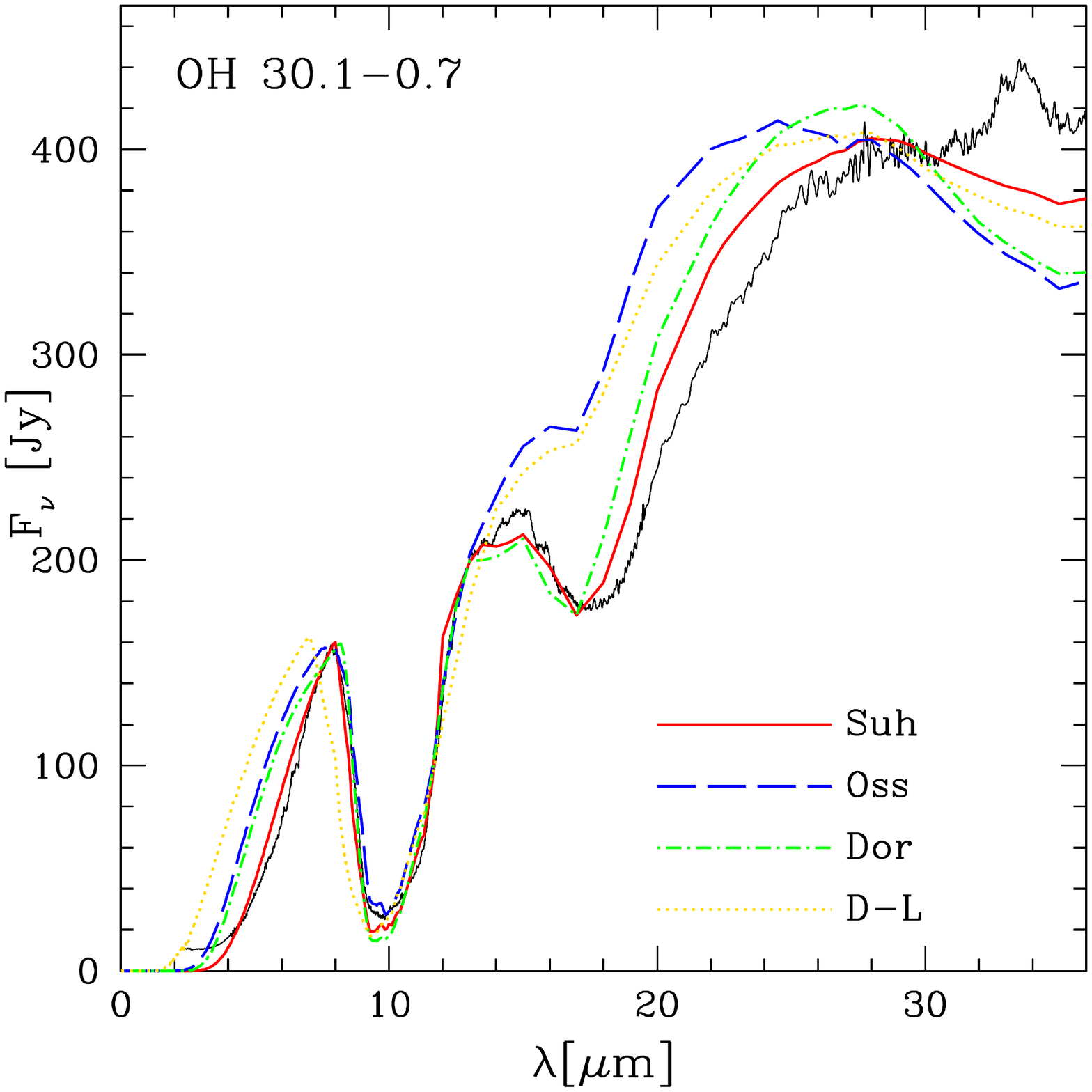}}
\end{minipage}
\vskip-50pt
\caption{\textit{ISO} spectrum of OH 30.1-0.7 (solid black line), together with four best-fit model spectra calculated adopting the following optical constants for silicates: \citet{suh99} (solid red line), \citet{oss92} (dashed blue line), \citet{dor95} (dash-dotted green line), and \citet{draine84} (dotted yellow line). The DUSTY inputs used for each model spectrum are reported in the text.}
\label{fopt}
\end{figure}

Regarding the dust mineralogy, the best agreement with the observations is found assuming the following mineralogy: a dominant contribution by silicates ($\sim$80\%, of which 5-10\% are under the form of crystalline), completed by smaller fractions of solid iron (10-20\%) and alumina dust ($\sim$5\%). For a given dust species, the choice of the optical constants on which the calculation of the extinction coefficients is based strongly affects the morphology of the synthetic SED. Silicates are generally the main dust components in the winds of M-type stars; therefore, they make a major contribution in this regard. We therefore explored different possibilities for this species, namely the optical constants by \citet[][D-L]{draine84}, \citet[][Oss]{oss92}, and \citet[][Dor]{dor95}, in order to identify which optical constants allow for the best fit of the observations. We further considered the optical constants presented by \citet{suh99} with the specific aim of fitting the observations of AGB stars. The results of this analysis are shown in Fig.~\ref{fopt}, which reports a comparison of four of our predicted SEDs with the \textit{ISO} spectrum of OH 30.1-0.7, taken as an example. For each of the aforementioned optical constants, we show our best-fit model, characterised by $\tau_{10}$=11 for Suh and Dor and $\tau_{10}$=14 for DL and Oss; and $T_{dust}$=1050K and a mineralogy which is dominated by silicates in all cases (75-90\%), with smaller percentages of alumina dust and solid iron.

The poor agreement between the model spectra calculated with D-L, Oss, and Dor and the observations makes evident the limitations of these optical constants when we tried to fit such obscured sources. The  main discrepancies are as follows: i) the continuum at $\lambda$<8$~\mu$m is not well reproduced and the flux in the spectral region at 20$~\mu$m<$\lambda$<30$~\mu$m is overestimated, deviating from the observations of a factor $\sim$2 in the former case and between $\sim$5-15\% in the latter, depending on the wavelength; ii) D-L and Oss SEDs do not reproduce the depth or the shape of the $\lambda$=18$~\mu$m absorption feature at all; and iii) the flux in the far-IR ($\lambda$>30$~\mu$m) is underestimated in all three cases by $\sim$15\% and the synthetic SEDs show a steeper spectral slope than observed.

Overall we may conclude that the Suh optical constants are the only ones that allow for the global shape of the observed spectrum to be reproduced, leading us to adopt these coefficients in the analysis of the stars presented this work. However, the \citet{suh99} data are empirical and in some way designed to reproduce observed spectra. Thus, there is a need to systematically consider a wide range of combinations of laboratory measurements and theoretical calculations in the future.

Once the best input parameters for DUSTY were identified, we were able to determine the mass-loss rates of these stars by running DUSTY in the alternative modality described in the previous section, thus finding values of $\dot{\rm M}$ in the 1-2$\times 10^{-4}~{\rm M}_{\odot}/$yr range.
 % Unfortunately, the poor reliability of the \textit{Gaia} distances for such obscured sources prevented us to estimate their luminosity, which  can be inferred by scaling the synthetic SED until matching the observed one, if the distance is known.
An additional outcome of the SED analysis would be the determination of the luminosity, 
%which can be done by scaling the synthetic SED until matching the observed one, 
 as long as the distance of the star is known. Unfortunately, this step was not possible in the case of the present work since we cross-matched our candidates with the latest data release of the satellite \textit{Gaia} \citep[Gaia DR3,][]{gaia16, gaia22}.\ However, given the faintness of such obscured sources in the optical regime, we found that only one of the stars, RAFGL 5379, is included in the Gaia DR3 catalogue. The quoted parallax for RAFGL 5379 is negative with large uncertainties, (-0.122$\pm$0.553) mas, preventing us from estimating its 
 distance\footnote{Positive parallaxes with relative errors below $\sim$20\% can be inverted to derive a distance \citep[e.g.][]{bailer15}. In all the other cases, a Bayesian approach can be used to infer distances \citep[e.g.][for an application to AGB stars]{bailer21, distAGB}. In our case, the signal-to-noise of the only Gaia DR3 parallax available is so low that we decided not to implement this, so as to not bias our results as to the choice of the adopted prior.} and consequently the  luminosity.

\begin{table}
\caption{Four sources analysed in this work with the isotopic carbon ratios derived by \citet{kate13} and the observed periods with the following references: 1-\citet{olivier01}; 2-\citet{engels15}; 3-\citet{suh02}; 4-\citet{langevelde90}; 5-\citet{martin22}; and 6-\citet{wolak13}.}
\label{table_info}      
\centering
%\addtolength{\leftskip}{-2cm}
%\addtolength{\leftskip}{-2cm}
\begin{tabular}{c c c c}    
\hline      
Source Name & $^{12}$C$/^{13}$C & P[days] & Ref. \\
\hline 
RAFGL 5379 & 27$\pm$11 & 1440 & 1 \\ 
OH 26.5+0.6 & 30$\pm$16 & 1591 & 2 \\
 &  & 1559$\pm$7 & 3 \\
 &  & 1589$\pm$42 & 4 \\
OH 30.1-0.7 & 4$\pm$1 & 2171 & 1 \\
 &  & 1952$\pm$46 & 5 \\
 &  & 2013$\pm$243 & 4 \\
OH 127.8-0.0 & 2$\pm$1 & 1590 & 2  \\
 &  & 1600 & 6 \\
\hline
\end{tabular}
\end{table}

\section{The AGB evolution and dust formation of intermediate mass stars}
\label{agbev}

Results from stellar evolution and dust formation modelling \citep{ventura14, ventura18} lead us to consider the stars of intermediate mass (${\rm M} \geq 4~{\rm M}_{\odot}$) as the best candidates in the interpretation of the sources analysed here, based on the fact that they are expected to produce the highest amounts of silicates in their winds.
%of initial mass ${\rm M} \geq 4~{\rm M}_{\odot}$\footnote{Also low-mass AGB stars (${\rm M} \leq 1.5~{\rm M}_{\odot}$) evolving through the phases before becoming carbon stars produce non-negligible amounts of silicates; however they are not expected to produce such large amount of dust, reaching maximun values of optical depth of $\sim$0.1 \citep{flavia15a,ester20}.}. 
%since  low-mass AGB stars (${\rm M} \leq 1.5~{\rm M}_{\odot}$) that produce silicates before becoming C-stars are not expected to produce such large amounts of dust \citep{flavia15a}.
The evolution of intermediate mass AGB stars of solar metallicity was studied by \citet{ventura18}, who discuss the main evolutionary properties of these stars, the efficiency of the dust production mechanism in their circumstellar envelope, and the uncertainties associated with the description of their AGB evolution, primarily connected to the still limited knowledge on convective instability and mass-loss mechanisms.

%The evolution of ${\rm M} \geq 4~{\rm M}_{\odot}$ stars is characterized by the ignition of HBB at the base of the convective envelope \citep{sackmann91}. This mechanism is due to the partial overlapping of the convective mantle with the H-burning shell, which triggers the activation of an advanced proton capture nucleosynthesis in the inner regions of the surface convective zone. The activation of HBB drives the evolution of the surface chemical composition of the star, which will reflect the equilibria of the CNO nucleosyntyhesis. HBB also affects the luminosity of the star, which after the beginning of HBB grows faster than in the earlier AGB phases \citep{blocker91, ventura05}. The ignition of HBB requires core masses above $\sim 0.8~{\rm M}_{\odot}$ \citep{ventura13}, which is the reason why it is experienced only by stars of intermediate mass.

\begin{figure*}
\begin{minipage}{0.48\textwidth}
\resizebox{1.\hsize}{!}{\includegraphics{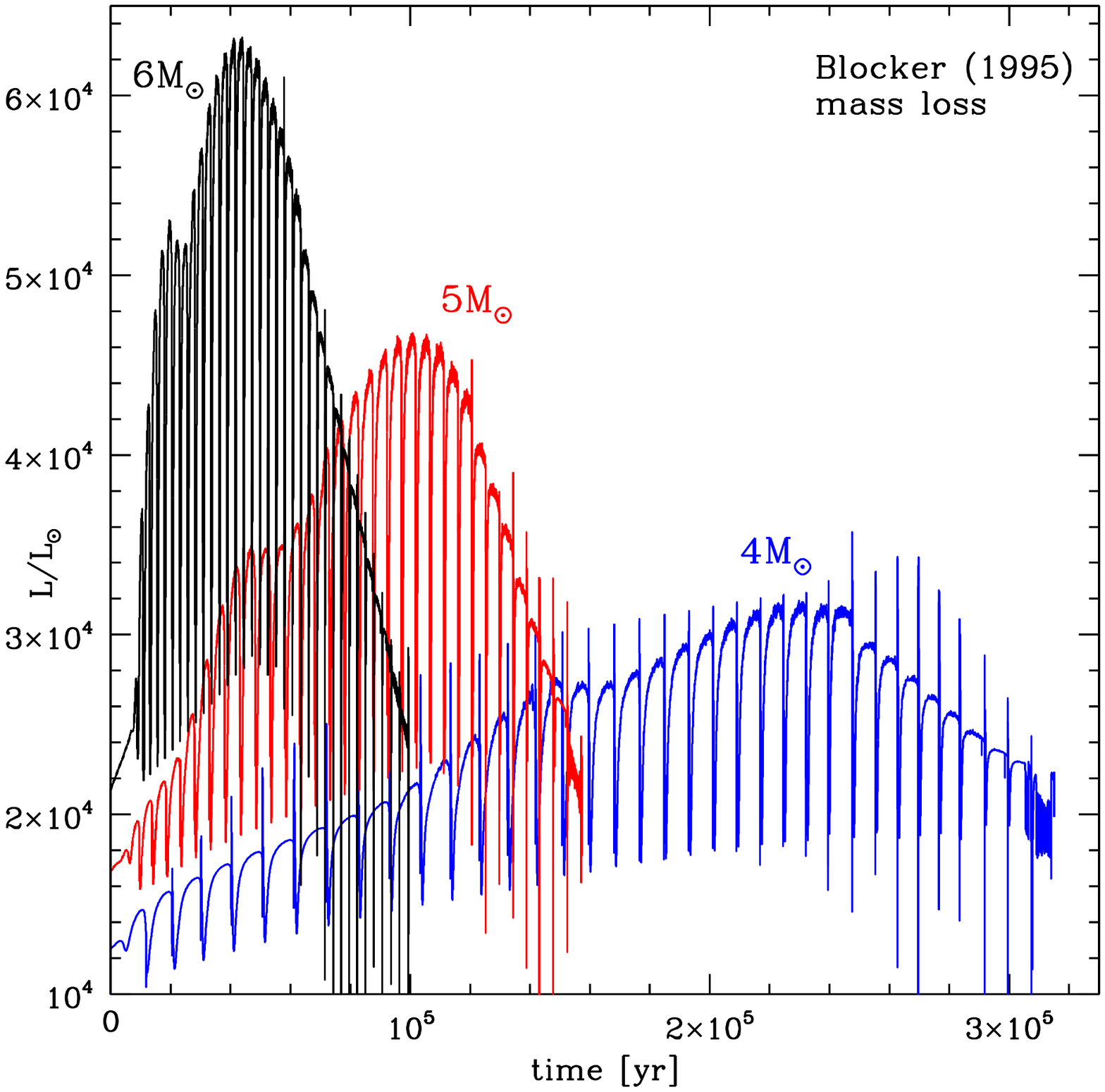}}
\end{minipage}
\begin{minipage}{0.48\textwidth}
\resizebox{1.\hsize}{!}{\includegraphics{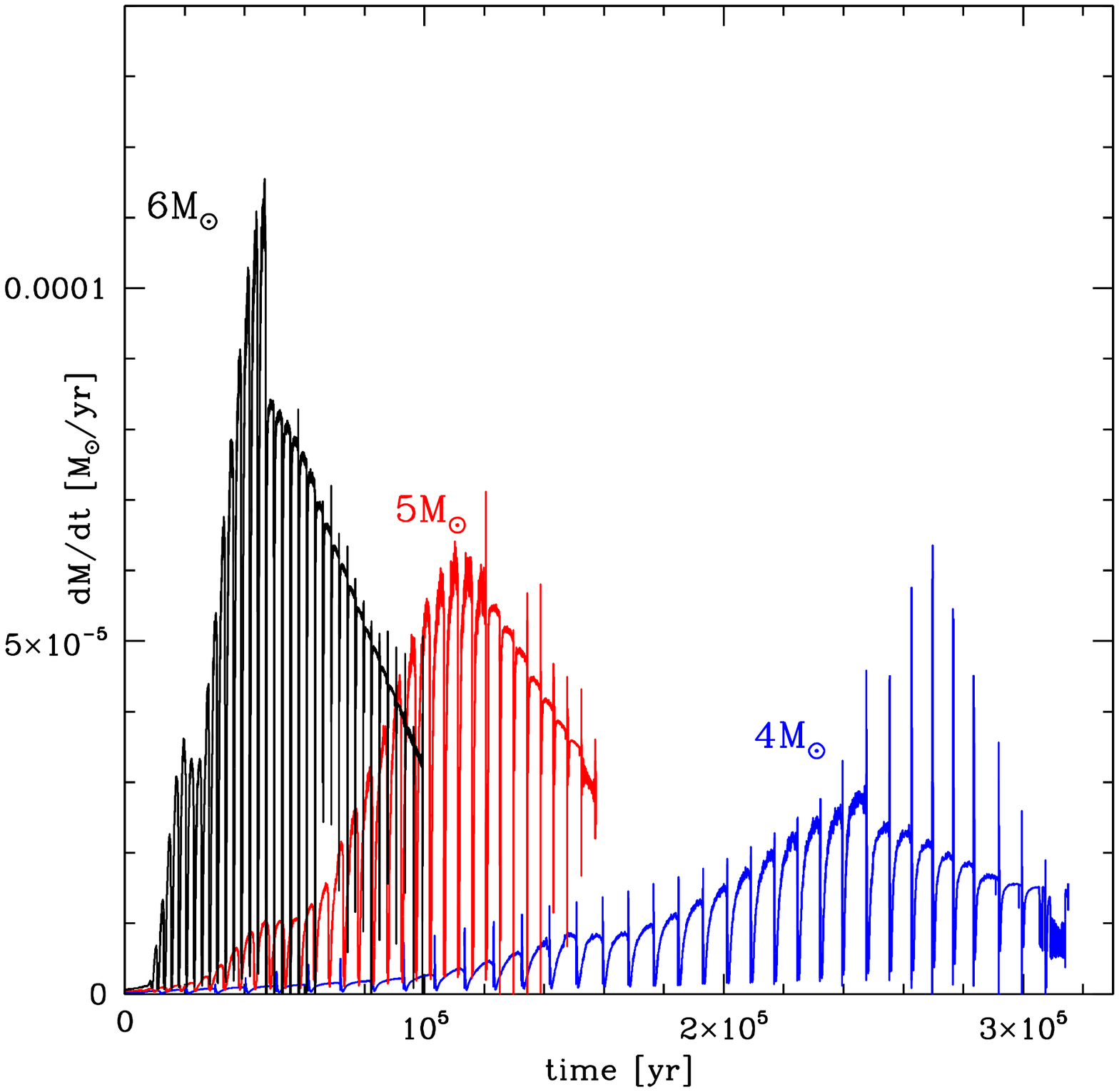}}
\end{minipage}
\vskip-80pt
\begin{minipage}{0.48\textwidth}
\resizebox{1.\hsize}{!}{\includegraphics{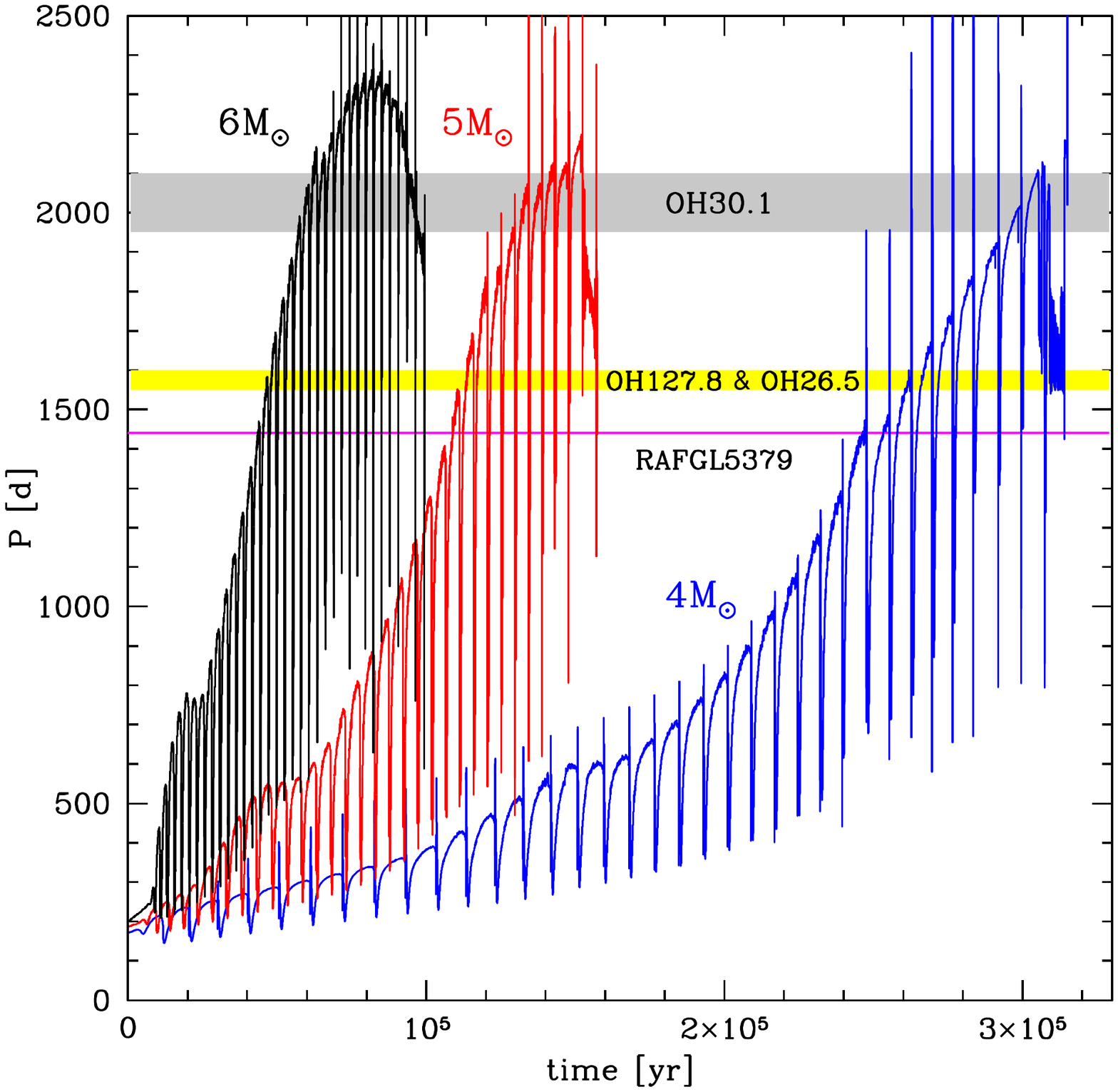}}
\end{minipage}
\begin{minipage}{0.48\textwidth}
\resizebox{1.\hsize}{!}{\includegraphics{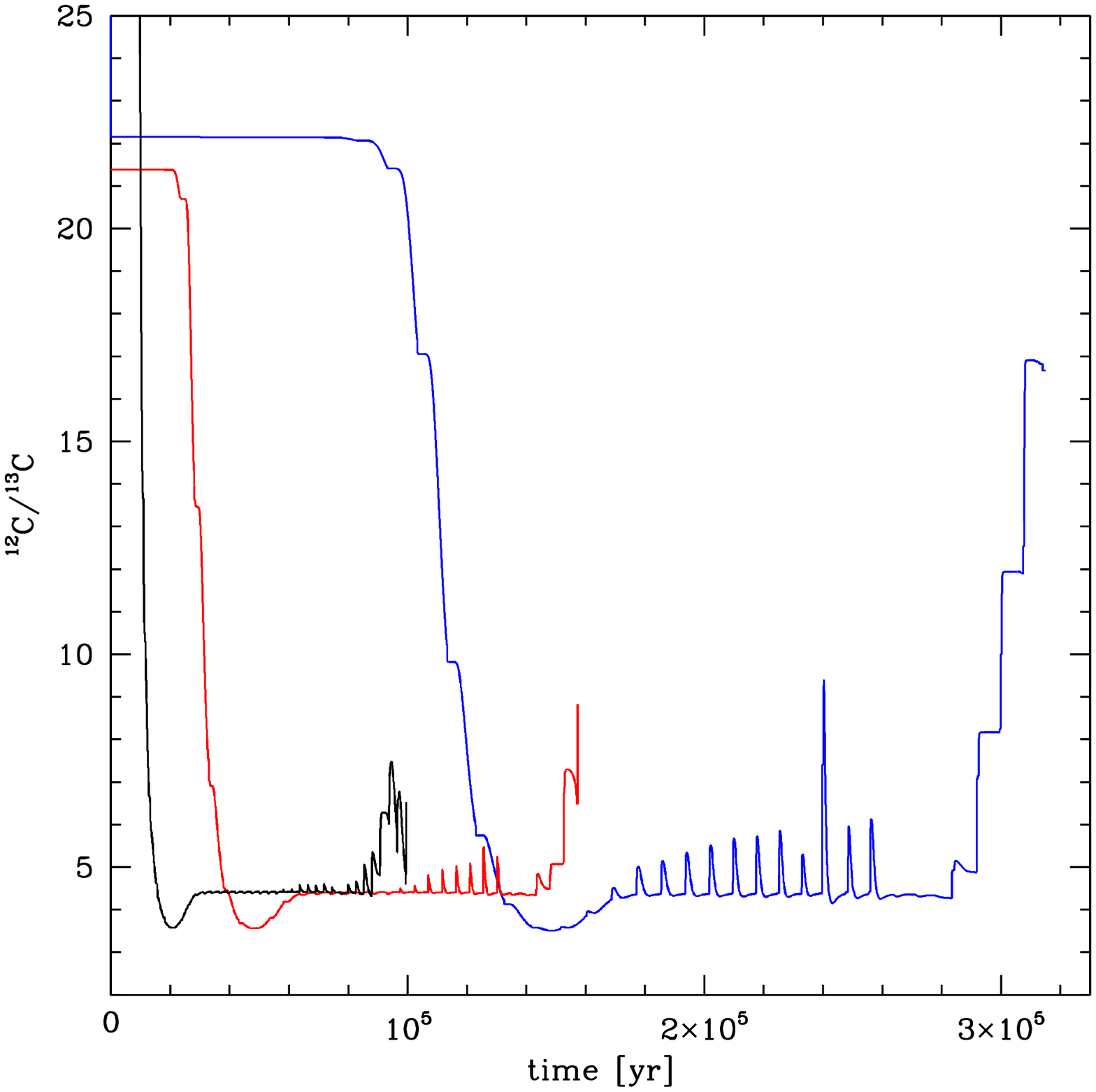}}
\end{minipage}
\vskip-50pt
\caption{AGB evolution of the main physical and chemical properties
of the model stars with a solar metallicity with an initial mass of $4~{\rm M}_{\odot}$ (blue lines),
$5~{\rm M}_{\odot}$ (red), and $6~{\rm M}_{\odot}$ (black), as a
function of time, counted since the beginning of the AGB phase.
The behaviour of luminosity (top left panel), mass-loss rate
(top right), pulsation period (bottom left), surface
$^{12}$C$/^{13}$C (bottom right) are shown. The shadings in the bottom left panel indicate the range of values of the pulsation periods taken from the literature of: OH 30.1-0.7 (grey), 127.8-0.0 and OH 26.5+0.6 (yellow), and RAFGL 5379 (magenta).
}
\label{evol}
\end{figure*}

%The evolution of intermediate mass AGB stars of solar metallicity was studied by \citet{ventura18}, who discussed the main evolutionary properties of these stars, the efficiency of the dust production mechanism in their circumstellar envelope, and the uncertainties associated to the description of their AGB evolution, primarily connected to the still limited knowledge convective instability and mass-loss mechanisms.
The evolution of ${\rm M} \geq 4~{\rm M}_{\odot}$ stars is characterised by the ignition of hot bottom burning (HBB) at the base of the convective envelope \citep{sackmann91}. This mechanism is due to the partial overlapping of the convective mantle with the H-burning shell, which triggers the activation of advanced proton-capture nucleosynthesis in the inner regions of the surface convective zone. The activation of HBB drives the evolution of the surface chemical composition of the star, which reflects the equilibria of the CNO nucleosyntyhesis. HBB also affects the luminosity of the star, which after the beginning of HBB grows faster than in the earlier AGB phases \citep{blocker91, ventura05}. The ignition of HBB requires core masses above $\sim 0.8~{\rm M}_{\odot}$ \citep{ventura13}, which is the reason why it is experienced only by stars of intermediate mass.

Fig.~\ref{evol} shows the time variation of the main physical quantities
of ${\rm M} \geq 4~{\rm M}_{\odot}$ model stars of solar metallicity presented in
\citet{ventura18} during the AGB phase. For clarity concerns, we show only the results of 
stars with an initial mass of $4, 5,$ and $ 6~{\rm M}_{\odot}$. In looking at the top left panel 
of Fig.~\ref{evol}, one can recognise the typical behaviour of the luminosity of these stars \citep{ventura22}: 
the initial phase, during which the luminosity increases owing to the growth of the core
mass, is followed by a second phase, during which the luminosity decreases, as HBB
is gradually extinguished. The peak luminosity of the stars increases with the
initial mass as the core mass of the stars during the AGB phase is higher, the larger
the initial mass \citep{ventura13, karakas14}. This affects the duration of the 
AGB phase, which is anti-correlated
with the initial mass: the range of the timescales of the AGB evolution increases from
$\sim 2\times 10^5$ yr, for $8~{\rm M}_{\odot}$ model stars, to $\sim 3\times 10^6$ yr, for 
${\rm M} = 4~{\rm M}_{\odot}$ \citep{ventura18}.

The mass-loss rate experienced by the stars, shown in the top right panel of Fig.~\ref{evol},
scales approximately with the luminosity. This is due to the tight relationship between
the luminosity and the mass-loss rate in the \citet{blocker95} treatment (see equation 1).
We note, in particular, the decrease in the mass-loss rate characterising the very final AGB phases, 
during which $\dot{\rm M} \sim 10^{-5}~{\rm M}_{\odot}/$yr. The maximum mass-loss rate
 also changes with the initial mass: specifically for the model stars
shown in Fig.~\ref{evol}, the peak values of $\dot{\rm M}$ are $3\times 10^{-5}~{\rm M}_{\odot}/$yr,
$7\times 10^{-5}~{\rm M}_{\odot}/$yr, and $10^{-4}~{\rm M}_{\odot}/$yr for the $4, 5,$ and $ 6~{\rm M}_{\odot}$
model stars, respectively. 

The variation of the pulsation period of the star, shown in the bottom left panel
of Fig.~\ref{evol}, is not correlated with luminosity as much as the mass-loss 
rate is. This is because the star continues to expand even after the
luminosity peak is reached, a behaviour typical of stars surrounded by a
convective mantle that is progressively lost via stellar wind. The contraction
phase starts only at the very end of the AGB evolution, when the residual
mass of the envelope drops below $\sim 0.2~{\rm M}_{\odot}$, and 
the CNO cycle is no longer sufficiently efficient to support
the star on the energetic side. As shown in Fig.~\ref{evol}, the period
of these stars grows during the first part of the AGB
evolution, from a few hundred days (d) until exceeding 2000 d, then 
it decreases to $\sim 1500$ d during the very final AGB evolutionary stages.

To describe the variation of the surface chemical composition, we show the variation of the 
$^{12}$C$/^{13}$C carbon ratio in
the bottom right panel of the same figure. The onset of HBB is clearly visible 
in the drop of the carbon ratio taking place during the first AGB phases, 
which continues until the equilibrium value $^{12}$C$/^{13}$C$\sim 4$ is
reached: This demonstrates the full effect of HBB in
modifying the surface chemistry. In Fig.~\ref{evol} one can recognise the effects
of a third dredge-up (TDU) in the fast increase in $^{12}$C$/^{13}$C that follows each thermal 
pulse, and in the increase in the carbon ratio that characterises the final
AGB phases, after HBB was turned off. Further effects of the ignition of
HBB are the synthesis of nitrogen and sodium, which during the AGB lifetime 
increase by a factor of $\sim 5$ and $\sim 3$, respectively, and the destruction 
of the surface $^{18}$O, the most fragile among the oxygen isotopes.
Intermediate mass stars evolve through the so-called lithium-rich phase,
during which large quantities of lithium are synthesized by the \citet{cameron}
mechanism. The duration of the lithium-rich phase depends on the rate at which 
the surface $^3$He is consumed, and it accounts for $\sim 50\%$,
$\sim 40\%$, and $\sim 30\%$ of the AGB phase, for the model stars of initial mass
$4~{\rm M}_{\odot}$, $5~{\rm M}_{\odot}$, and $6~{\rm M}_{\odot}$, respectively.
The HBB experienced by solar-metallicity stars is soft, thus the
temperatures reached by the innermost regions of the convective envelope are
not sufficiently hot to activate more advanced nucleosynthesis, typical of
lower metallicity stars \citep{flavia18b}: Neither the depletion of $^{16}$O 
and $^{24}$Mg, nor the increase in the surface abundances of aluminium and 
silicon takes place in the model stars examined here.

Regarding dust, the study by \citet{ventura18} suggests that most of the dust 
that formed in the circumstellar envelope of massive AGBs is composed
of silicates ($70\%-80\%$), with alumina dust and solid iron making up the
remaining $20\%-30\%$. The fraction of gaseous silicon that condensed into 
dust is in the $20\%-30\%$ range, whereas the fraction of aluminium that condensed
into Al$_2$O$_3$ ranges from $\sim 30\%$ (initial and final AGB phases)
to $\sim 80\%$ (in correspondence of the luminosity peak). The typical
gas-to-dust ratio found in \citet{ventura18} is in the $500-1000$ range.

\begin{figure*}
\begin{minipage}{0.32\textwidth}
\resizebox{1.\hsize}{!}{\includegraphics{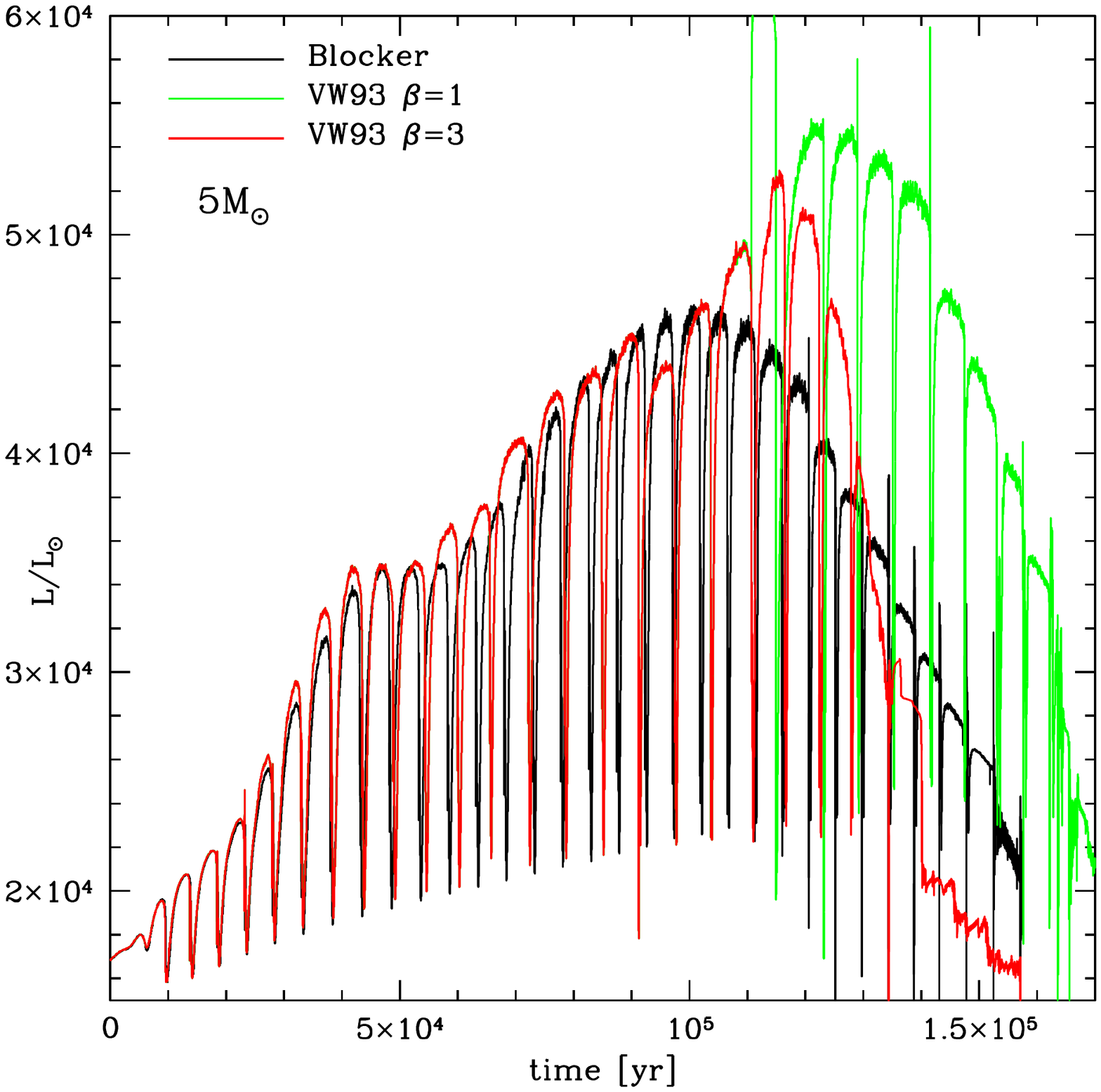}}
\end{minipage}
\begin{minipage}{0.32\textwidth}
\resizebox{1.\hsize}{!}{\includegraphics{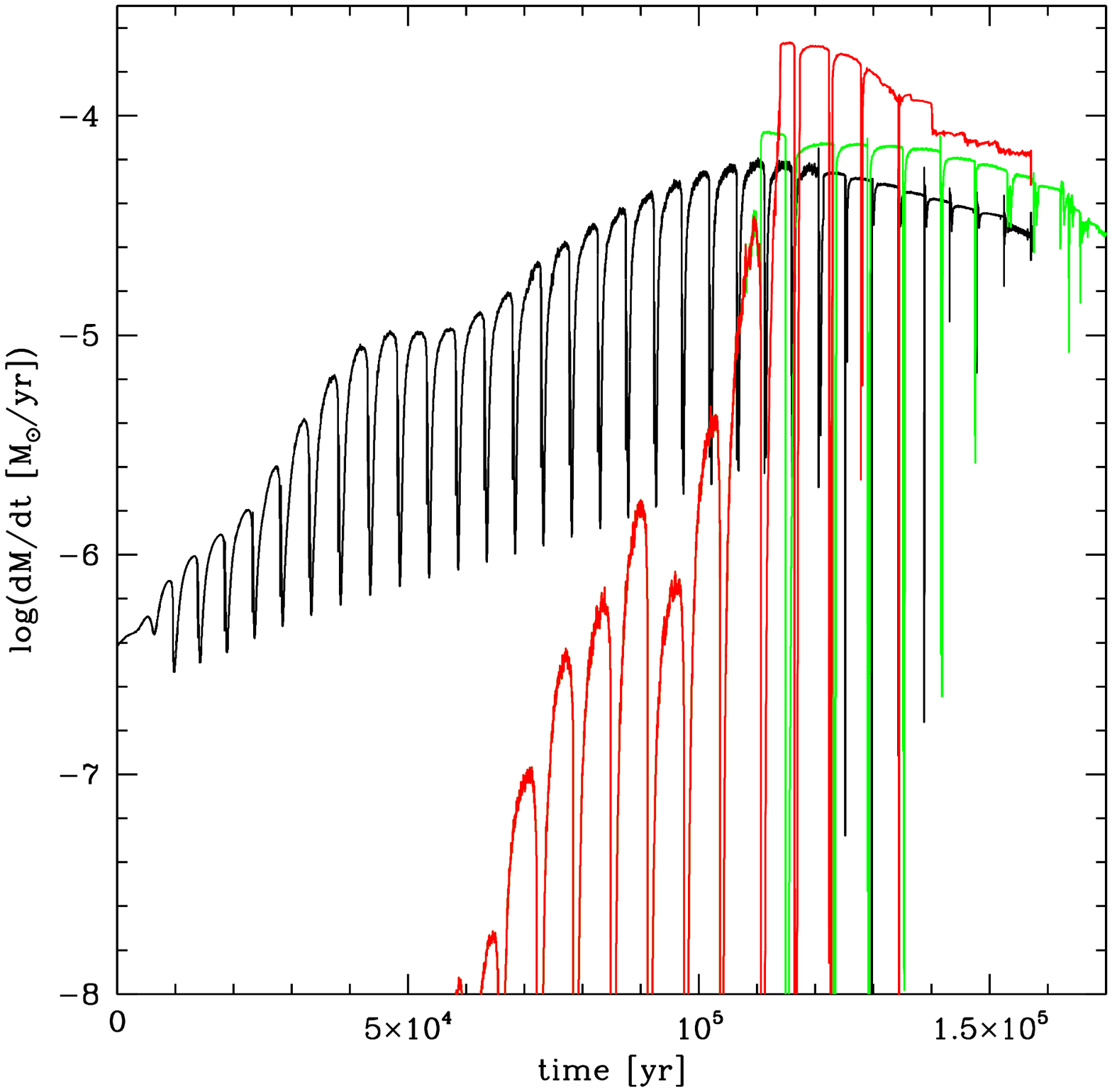}}
\end{minipage}
\begin{minipage}{0.32\textwidth}
\resizebox{1.\hsize}{!}{\includegraphics{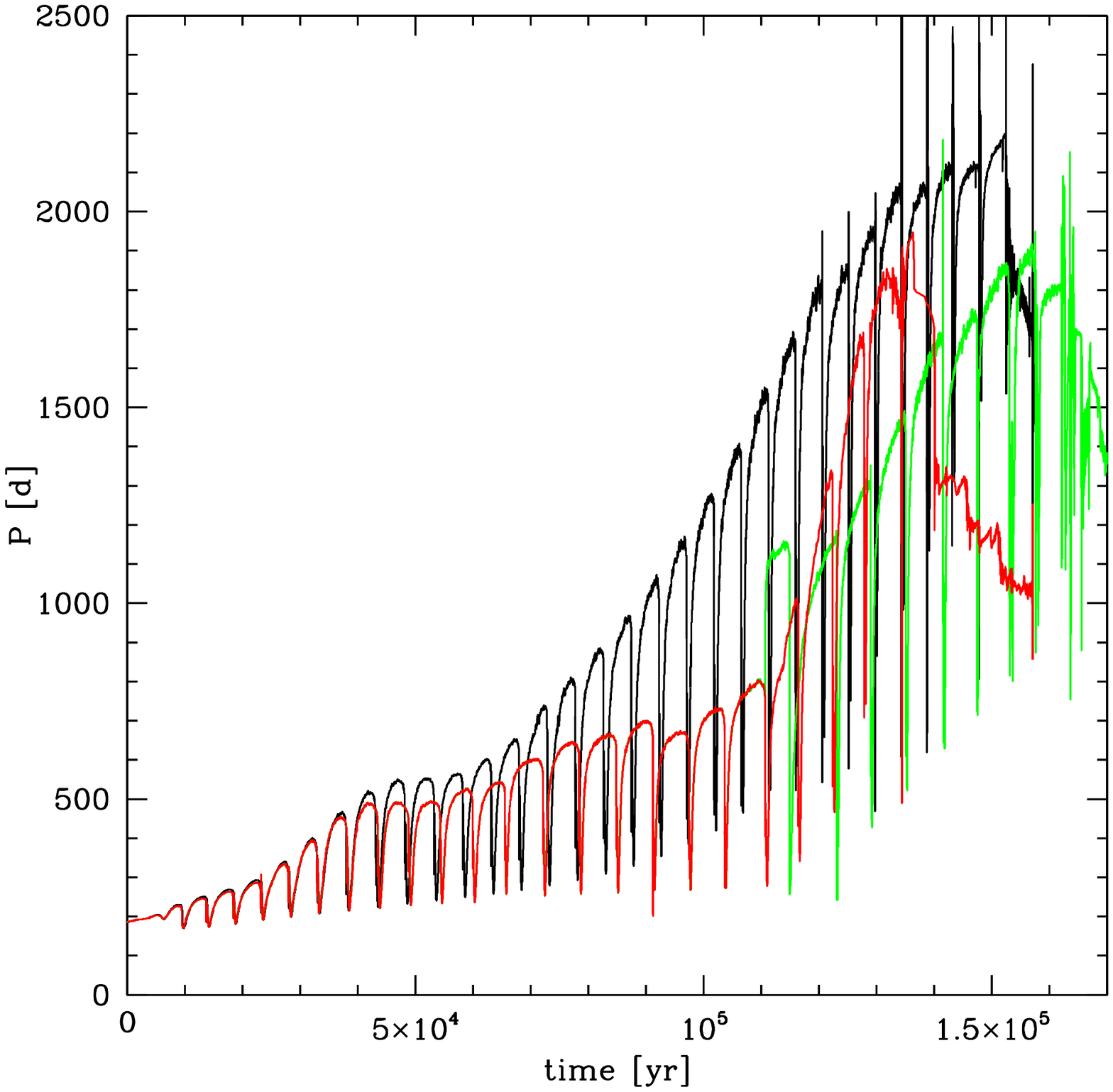}}
\end{minipage}
\vskip-40pt
\caption{Time variation during the AGB phase of the luminosity
(left panel), mass-loss rate (middle), and pulsation period
(right) of solar metallicity, model stars of initial mass 
$5~{\rm M}_{\odot}$, in which mass loss was modelled according
to \citet{blocker95} (black lines) and to \citet{vw93},
where the scattering parameter $\beta$ was set to 1
(green lines) and 3 (red).
}
\label{f5msun}
\end{figure*}

\section{The role of mass loss}
\label{mloss}
The empirical mass-loss treatment proposed by VW93 is based on an empirical formula
relating the mass-loss rate to the pulsation period. 
According to VW93 the mass-loss rate increases with the pulsation period, 
until reaching the super-wind phase, when the radiation-driven mass-loss 
rate is adopted, according to the expression 
$\dot{\rm M}=\beta {\rm L}/({\rm v}_{\rm exp}c)$, discussed in section 
\ref{models_evolution}.

The differences between the results obtained with the Blo95 and VW93
treatments applied to the modelling of a $5~{\rm M}_{\odot}$ star are shown in
Fig.~\ref{f5msun}. For the VW93 case, we considered $\beta=1$ and a further 
case in which we set $\beta=3$, starting from the maximum luminosity phase.

As shown in the middle panel of Fig.~\ref{f5msun}, the model star calculated 
with the Blo95 treatment experiences
higher mass-loss rates during the first part of the AGB evolution, thus
the envelope is expelled faster. This makes the duration of the AGB phase of VW93 model longer, which allows for higher growth of the core mass, and thus higher peak luminosities.
%This is the reason why the VW93 model stars reach higher peak luminosities. 
The more rapid consumption of the envelope
is also the reason why, in the $\beta=1$ case, the
luminosity reached is higher than for $\beta=3$. The Blo95 model star evolves 
at larger periods compared to the VW93 ones: This is once more due
to the higher mass loss experienced during the initial part of the AGB
phase, which makes the star reach a more expanded configuration, and hence
experience longer pulsation periods. The Blo95 and the VW93 
$\beta=1$ model stars experience similar peak mass-loss rates, of the order of
$10^{-4}~{\rm M}_{\odot}/$yr; on the other hand, in the VW93 $\beta=3$
case, the largest mass-loss rate experienced is $2\times 10^{-4}~{\rm M}_{\odot}/$yr.

These rates of mass loss are at odds with the results from dust formation in the winds of M-type stars in which stellar pulsation and associated shocks are properly considered \citep{bladh13, bladh19} and which rarely form outflows of more than $\dot{M} \sim 10^{-5}\,M_\sun$~yr$^{-1}$. However, these works are mostly based on smaller luminosities than those invoked here (L $\geq 40000~{\rm L}_{\odot}$). More detailed investigations of dust production in the present luminosity domain is required before solid conclusions can be drawn in this regard.

\begin{figure*}
\begin{minipage}{0.48\textwidth}
\resizebox{1.\hsize}{!}{\includegraphics{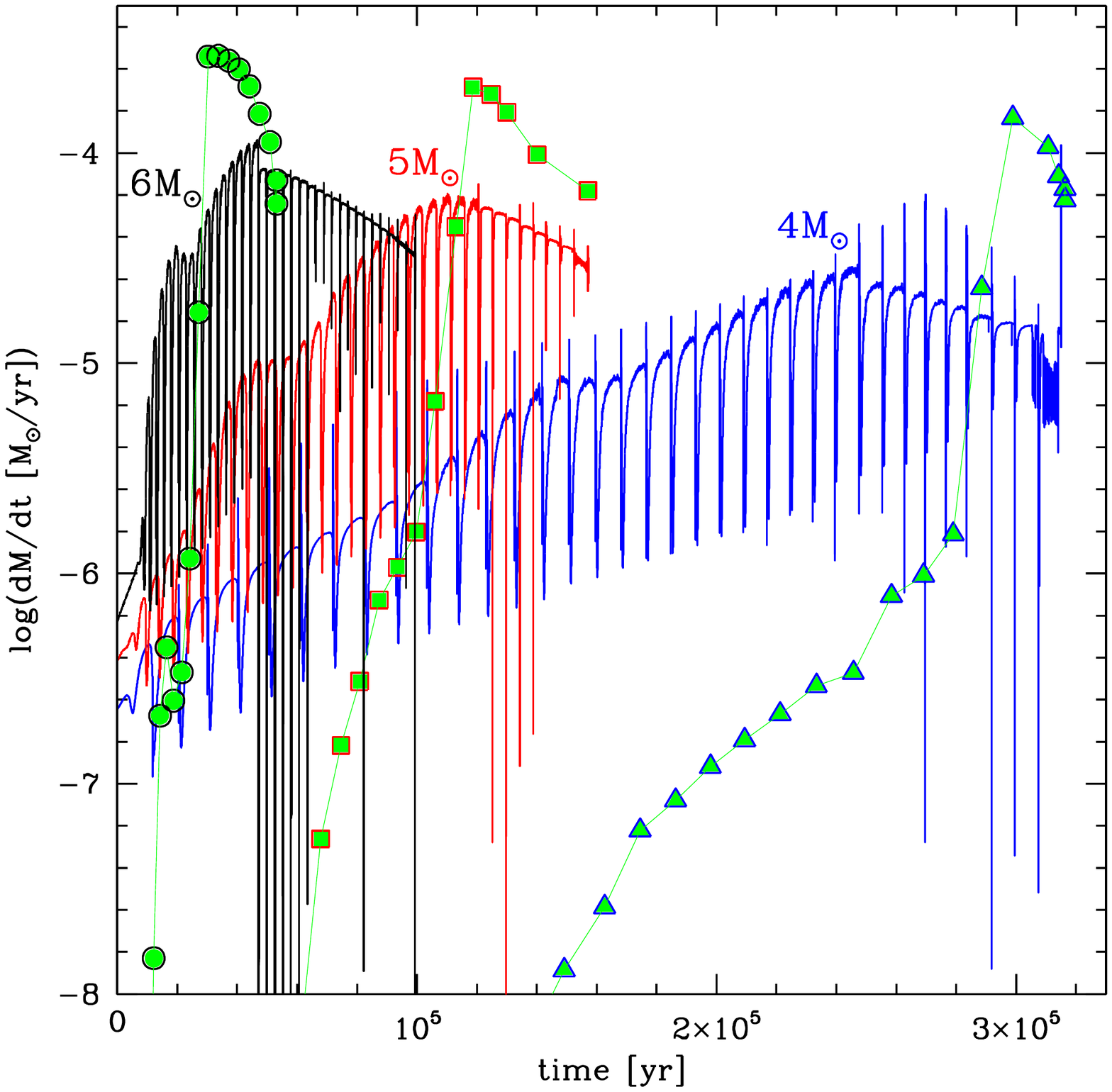}}
\end{minipage}
\begin{minipage}{0.48\textwidth}
\resizebox{1.\hsize}{!}{\includegraphics{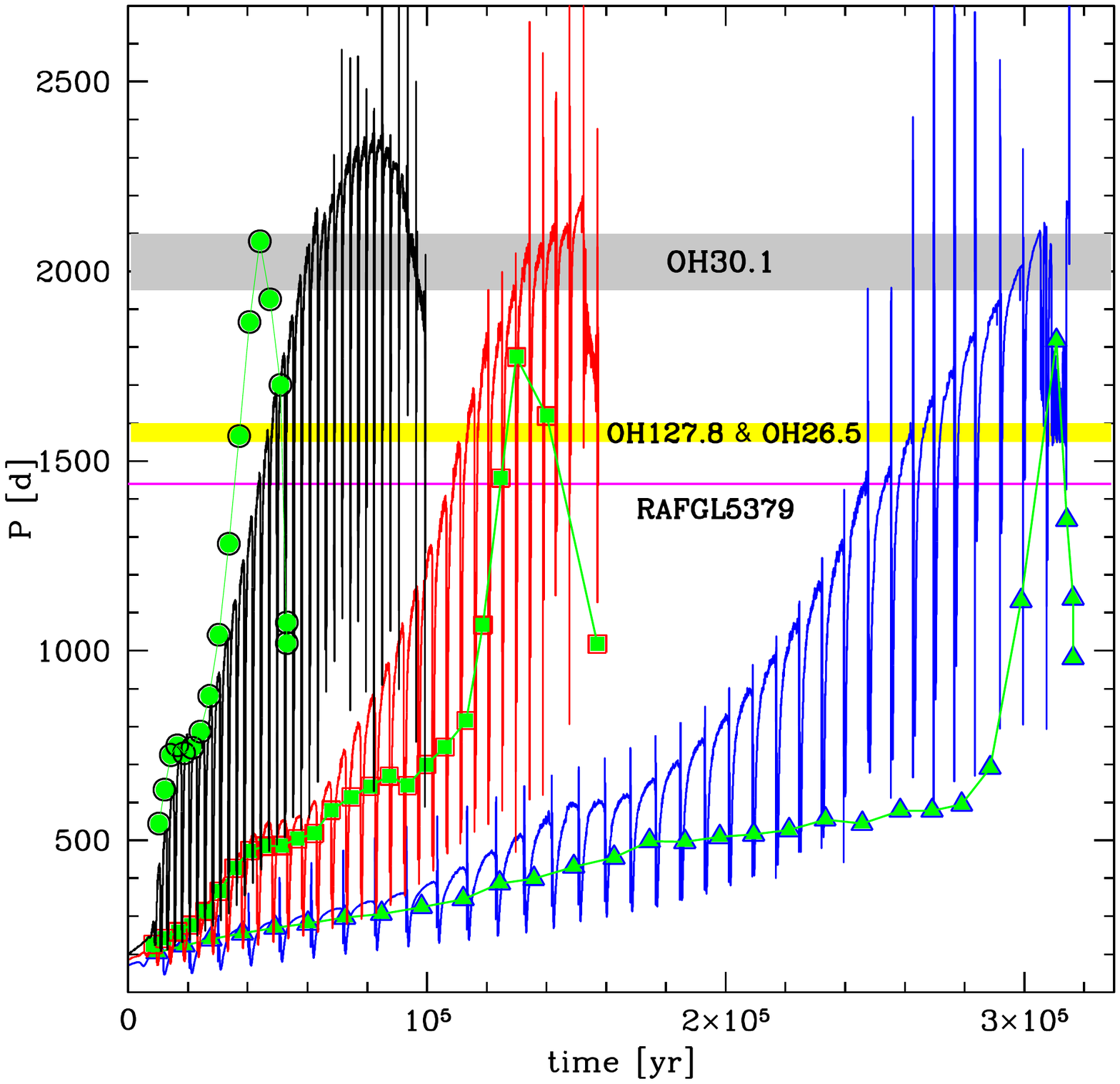}}
\end{minipage}
\vskip-80pt
\begin{minipage}{0.48\textwidth}
\resizebox{1.\hsize}{!}{\includegraphics{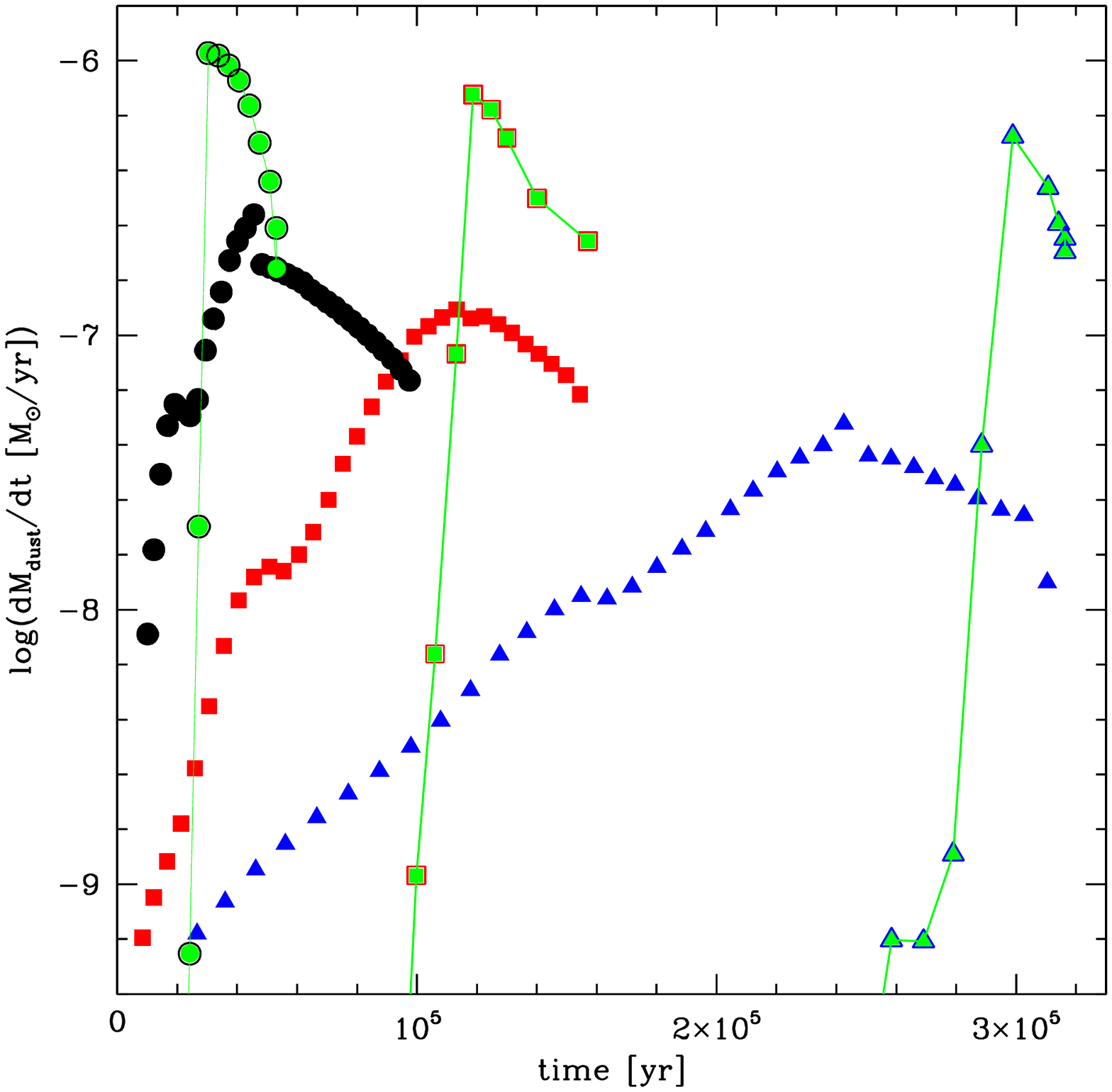}}
\end{minipage}
\begin{minipage}{0.48\textwidth}
\resizebox{1.\hsize}{!}{\includegraphics{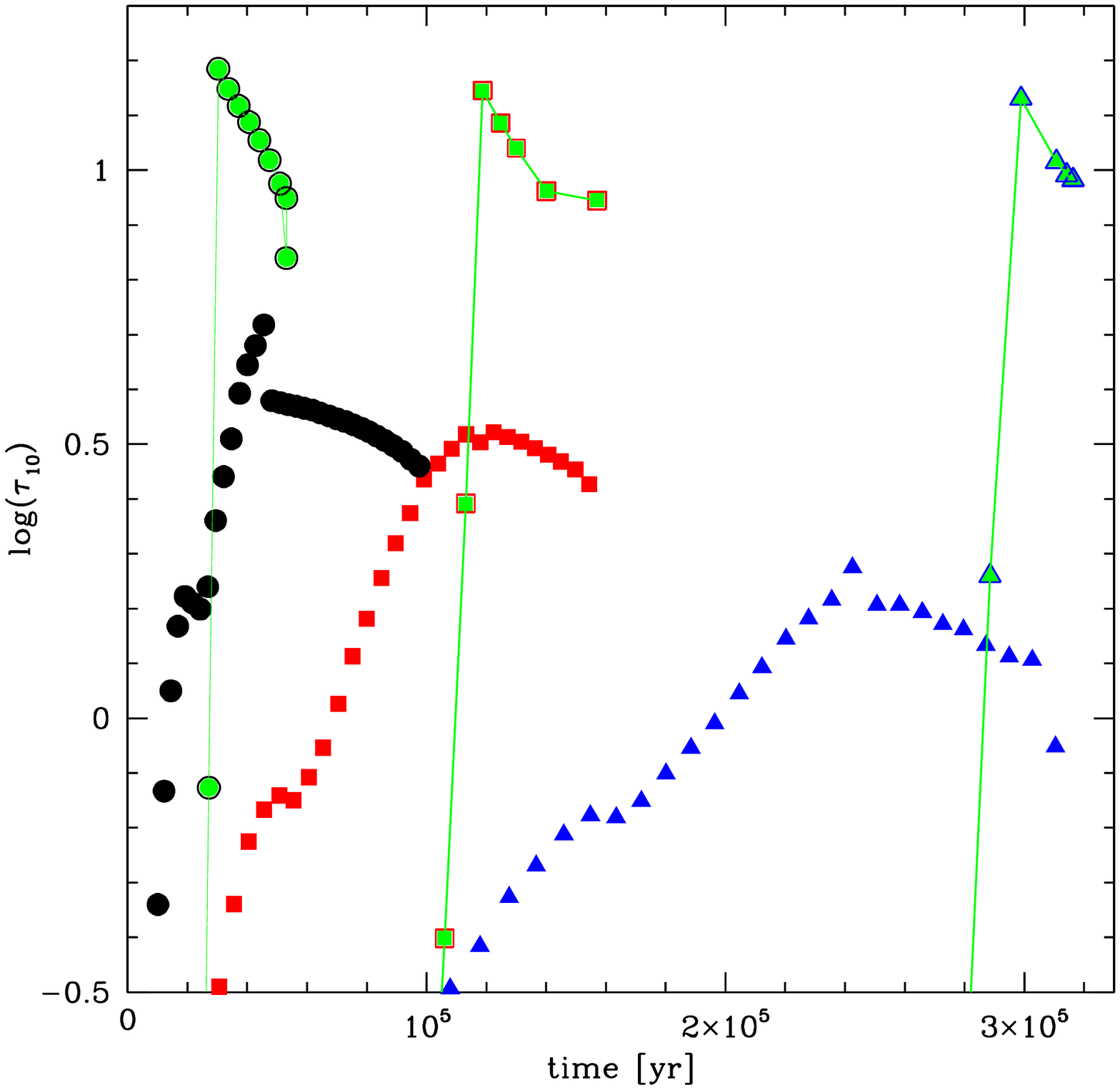}}
\end{minipage}
\vskip-50pt
\caption{Time variation of the mass-loss rate (top left panel), pulsation period 
(top right), dust production rate (bottom left), and optical depth $\tau_{10}$ 
(bottom right) of model stars
with an initial mass of $4~{\rm M}_{\odot}$ (black lines and circles),
$5~{\rm M}_{\odot}$ (red lines and squares), and $6~{\rm M}_{\odot}$
(blue lines and triangles), calculated with the mass-loss prescription
by \citet{blocker95}. The points in the bottom panels refer to the
values in the middle of each inter-pulse phase. Green points 
refer to the inter-pulse phases, and indicate results from modelling based on the VW93 description of mass loss, where the value $\beta=3$ was
considered for the scattering parameter. The shadings in the top right panel are the same as in Fig.~\ref{evol}.
}
\label{fall}
\end{figure*}

Significant differences among the mass-loss rates of the various model stars 
are found during the very final phases, when the Blo95 model star 
experiences $\dot{\rm M}$ values being significantly smaller than the W93 counterparts,
owing to the previously discussed sensitivity of the Blo95 treatment on
the luminosity of the star.
The full comparison among the results obtained by different descriptions of
mass loss is shown in Fig.~\ref{fall}, where the time variation of the
mass-loss rate, pulsation period, dust production rate, and optical
depth at $10~\mu$m of intermediate mass AGBs are shown. We restrict the comparison 
to the Blo95 and VW93 model stars, calculated with $\beta=3$. 

The evolutionary timescales of the $4~{\rm M}_{\odot}$ and $5~{\rm M}_{\odot}$
model stars obtained by modelling mass loss with Blo95 and VW93 are similar: The 
depletion of the envelope during the first part of the AGB phase is faster 
in the Blo95 case, but this is counterbalanced by the higher luminosity attained 
by the VW93 models, which renders the timescales of the final phases shorter. 
In the $6~{\rm M}_{\odot}$ case,the differences between the luminosity of the 
Blo95 and VW93 models arise from the initial AGB phases, making the 
duration of the AGB phase of the VW93 model shorter.

Significant differences are found in the mass-loss rates experienced.
The VW93 rates are orders of magnitude smaller than Blo95
during the initial AGB phase, until the peak luminosity was reached.
When the super-wind takes over the peak $\dot{\rm M}$ values
of the VW93 model, stars are $\sim 3$ times higher than Blo95.

These differences in the mass-loss rate are the result of differences in the dust production rate of the stars,
for what concerns both the general behaviour across the AGB lifetime and
the largest value reached. As shown in the bottom left panel of Fig.~\ref{fall},
in the VW93 case, dust production is negligible during the first part
of the AGB phase, then dust is produced at rates of the order of 
$5\times 10^{-7}-10^{-6}~{\rm M}_{\odot}/$yr after the super-wind takes over.
Almost all the dust released by the stars is produced during the last four to five inter-pulse
phases. Conversely, the behaviour of $\dot{\rm M}_{\rm dust}$ with the AGB time
is smoother in the Blo95 case, and the largest rates
reached are $\sim 10^{-7}~{\rm M}_{\odot}/$yr. The gas-to-dust ratio
during the phases when the dust production rate is largest is $\sim 300$
in the VW93 case, which has yet to be compared to the minimum value of $\sim 500$ found 
by \citet{ventura18}.

The expected IR excess of the stars, here represented by $\tau_{10}$, 
is tightly connected to the dust production rate. This can be seen by comparing
the behaviour of $\dot{\rm M}_{\rm dust}$ with the
time variation of $\tau_{10}$, shown in the bottom right panel of 
Fig.~\ref{fall}. We note, in particular, the significant difference in the
largest $\tau_{10}$ values, attained during the final AGB phases: In the
VW93 model, star $\tau_{10}$ is in the $7-15$ range; whereas, in the Blo95 case,
we find $\tau_{10}<3$.

\section{The characterisation of the sample sources}
\label{sources}
Table~\ref{table_info} lists the pulsation periods and the isotopic carbon ratios of the four stars in the present study.
The sources are characterised by periods in the $1440$ d $<P<2000$ d range, as well as optical depths derived from SED fitting (see section~\ref{sed_fitting}), $8<\tau_{10}<16$. These results are not consistent with the outcome of modelling based on \citet{blocker95}, thus we subsequently focus on the results based on the VW93 treatment of mass loss.

\subsection{Stars experiencing hot bottom burning}
OH 127.8-0.0 and OH 30.1-0.7 exhibit isotopic carbon ratios $^{12}$C$/^{13}$C$=2\pm1$ and $^{12}$C$/^{13}$C$=4\pm1$, respectively, in agreement with the values expected when CNO nucleosynthesis reaches equilibrium; the periods from the literature are $\sim1600$ d for OH 127.8-0.0 and $\sim2000$ d for OH 30.1-0.7, whereas the optical depths
derived from SED fitting are $\tau_{10}=11$ (OH 30.1-0.7) and $\tau_{10}=13$
(OH 127.8-0.0). Based on the results shown in Fig.~\ref{fall}, we conclude that these sources are currently evolving through the AGB phases during which the dust production (hence mass-loss) rate are close to the maximum values, and they are currently experiencing HBB.
Unfortunately, as can be seen in Fig.~\ref{fall}, this information proves insufficient to assess the mass of the progenitor (hence the formation epoch) for OH 30.1-0.7 since all the model stars of initial mass in the $4-6~{\rm M}_{\odot}$ range evolve through phases characterised by the quantities given above. The accurate determination of the distance would be crucial to this scope, as the luminosity at which the star is expected
to evolve at the periods and optical depths given above is 
$\sim 3.5\times 10^4~{\rm L}_{\odot}$, $\sim 5\times 10^4~{\rm L}_{\odot}$, or $\sim 7\times 10^4~{\rm L}_{\odot}$, according to whether the initial mass of the star is $4~{\rm M}_{\odot}$, $5~{\rm M}_{\odot}$, or $6~{\rm M}_{\odot}$, respectively. Inversely, if the luminosity is known, it is possible to give an estimation of the distance of the star\footnote{This can be done by fixing the luminosity and shifting the synthetic SED until it matches the observed spectrum.}. If our understanding is correct, we would expect the following distances depending on luminosity: $\sim$3.5kpc ($\sim 3.5\times 10^4~{\rm L}_{\odot}$), $\sim$4.2kpc ($\sim 5\times 10^4~{\rm L}_{\odot}$), and $\sim$5kpc ($\sim 7\times 10^4~{\rm L}_{\odot}$).

On the other hand, the lower pulsation period of 127.8-0.0 
($\sim1600$d)  indicates that this star cannot descend from a 
progenitor with an initial mass $\sim$4$~{\rm M}_{\odot}$ because such 
periods correspond, in this case, to very low values of mass-loss rate (<$\sim10^{-6}~{\rm M}_{\odot}/$yr) and the theoretical optical 
depths (see the bottom right panel of Fig.~\ref{fall}) would be too low in comparison to those found by SED fitting. Therefore, we can 
conclude that 127.8-0.0 descends from stars of initial mass 
M>4$~{\rm M}_{\odot}$, and we expect luminosities depending on the 
mass of the progenitor, which are $\sim 5\times 10^4~{\rm 
L}_{\odot}$ (M=$4~{\rm M}_{\odot}$) and $\sim 7\times 10^4~{\rm 
L}_{\odot}$ (M=$6~{\rm M}_{\odot}$). These would correspond to 
distances of $\sim$5kpc and $\sim$6kpc, respectively.
%Unfortunately, as can be seen in Fig.~\ref{fall}, these information prove not sufficient the assess the mass of the progenitors (hence the formation epoch), since all the model stars of initial mass in the $4-6~{\rm M}_{\odot}$ range evolve through evolutionary phases characterized by the quantities given above. The accurate determination of the distance would be crucial to this scope, as the luminosity at which the stars are expected
%to evolve at the periods and optical depths given above is $\sim 3.5\times 10^4~{\rm L}_{\odot}$, $\sim 5\times 10^4~{\rm L}_{\odot}$, $\sim 7\times 10^4~{\rm L}_{\odot}$, according to whether the initial mass of the star is $4~{\rm M}_{\odot}$, $5~{\rm M}_{\odot}$ and $6~{\rm M}_{\odot}$, respectively.
The above discussed interpretation given for OH 127.8-0.0 and OH 30.1-0.7 is consistent with the mass-loss rates derived from SED fitting, which are in the $1.5-2.5\times 10^{-4}~{\rm M}_{\odot}/$yr range\footnote{The mass-loss rate derived from
 SED fitting via the DUSTY code depends on the luminosity of the star \citep{nenkova99}. When considering the luminosities $\sim 3.5\times 10^4~{\rm L}_{\odot}$, $\sim 5\times 10^4~{\rm L}_{\odot}$, and $\sim 7\times 10^4~{\rm L}_{\odot}$ given in
the text, we find that $\dot{\rm M}=1.5, 2.0, 2.5\times 10^{-4}~{\rm M}_{\odot}/$yr.}, also consistent with the results shown in the top left panel of Fig.~\ref{fall}.

\subsection{Stars evolving through the very late AGB phases}
The interpretation of RAFGL 5379 and OH 26.5+0.6 is more tricky because the
given $^{12}$C$/^{13}$C (27$\pm$11 and 30$\pm$16, respectively) are significantly higher than the equilibrium values.
The possibility that they are evolving through the initial AGB phase, when 
HBB has not yet started, can be ruled out since the periods would be much 
shorter than those measured (see Fig.~\ref{fall}). A further possibility is 
that these stars have just experienced a thermal pulse (TP), after which the action 
of TDU favours the increase in the surface $^{12}$C, and thus an off-equilibrium 
$^{12}$C$/^{13}$C; while this possibility cannot be ruled out, we consider it 
extremely unlikely since re-ignition of HBB favours a fast destruction of $^{12}$C 
in favour of $^{13}$C, thus the phase during which $^{12}$C$/^{13}$C is 
off-equilibrium lasts only for a very small fraction of the whole inter-pulse. 

We believe it is more plausible that these two stars are experiencing  late
AGB phases, during which HBB is turned off: as shown in the bottom right
panel of Fig.~\ref{evol}, the occurrence of a couple of TDU events
is sufficient to raise the carbon isotopic ratio above the equilibrium values.
This is in agreement with one of the possibilities invoked 
for RAFGL 5379 by \citet{devries14}.
The measured period of the two stars, $1440$ d for RAFGL 5379 and around $1600$ d 
for OH 26.5+0.6, further support this
hypothesis (see Fig.~\ref{evol}). The optical depth and the mass-loss rates
derived from SED fitting also support this picture.

A further confirmation that RAFGL 5379 and OH 26.5+0.6 experienced
HBB is given by the studies based on Herschel observations by \citet{kate13, kate15}, 
who could not detect any H$_2$O$_{18}$ line, despite the strong water emission
lines in H$_2$O$_{16}$ and H$_2$O$_{17}$.\ As discussed in section \ref{agbev},
intermediate mass stars of solar metallicity experience strong depletion of the 
surface $^{18}$O and negligible reduction of the surface $^{16}$O and $^{17}$O.

%Regarding OH 26.5+0.6, \citet{devries14}, based on SED modelling and on OH and CO 
%observations, suggested that the superwind phase, during which the dust responsible
%for the infrared excess was released, started not earlier than $\sim 2000$ yr ago.
%This is consistent with the possibility that this source descends from an
%intermediate mass progenitor. Indeed, considering that the inter-pulse periods of 
%the model stars explored here is $\sim 3000-5000$ yr, and that dust formation is expected 
%to stop after the ignition of each thermal pulse, it is reasonable to assume
%that this source is in an inter-pulse state, and that the last dust release 
%started $\sim 2000$ yr ago.

We note that unlike the two sources discussed above for which no or poor predictions 
regarding the luminosity was possible, for RAFGL 5379 and OH 26.5+0.6 we can claim values of
the order of $2\times 10^4~{\rm L}_{\odot}$, independently of the initial mass. Indeed, as can be seen in Fig.~\ref{fall}, the luminosities obtained by the stars during the very final AGB phase, after HBB was turned off, tend to converge towards the previously given value, substantially independently of the initial mass of the stars.
Considering this luminosity, the estimated distance in this case would be $\sim$1.2kpc for RAFGL 5379 and 
$\sim$1.8kpc for OH 26.5+0.6.
The mass-loss rate derived from SED fitting, when this luminosity is used,
is $10^{-4}~{\rm M}_{\odot}/$yr, consistent with the results shown in
the top left panel of Fig.~\ref{fall}.

\subsection{The binary possibility}
In a study based on results from ALMA observations of low-excitation rotational lines 
of $^{12}$CO, \citet{decin19} proposed that OH 26.5+0.6 and OH 30.1-07 are part of binary systems.
This conclusion is motivated by the detection of an incomplete ring-like pattern,
which is characteristic of a shell-like spiral, in turn connected to the orbital 
motion of the AGB star around the centre of mass of a binary system. On general grounds,
binarity offers
an interesting explanation to interpret the morphology of the gaseous emission of the circumstellar envelope extreme OH$/$IR stars.
%, alternative to the one proposed here. 

If the stars belong to binary systems, the large dust production rates would be related to
significant equatorial density enhancements, triggered by the gravitational attraction
of the companion star. In this case the largest single scattering mass-loss rates required
would be of the order of a few $10^{-5}~{\rm M}_{\odot}/$yr, a few times smaller than 
those needed in the case of single stars. 

On the statistical side, we do not see any
significant shortcomings from either possibilities. Binarity is found to be a 
rather common feature of intermediate mass stars, with percentages estimated to be
$30-40\%$. On the other hand, in the single star hypothesis, we note from the results
shown in Fig.~\ref{fall} that the time interval during which the stars experience 
a super-wind-like mass loss, with $\dot{\rm M} > 10^{-4}~{\rm M}_{\odot}/$yr, is 
20-40 kyr, which represents $10-50\%$ of the entire AGB lifetime, with the percentage 
being higher the larger the initial mass of the star. We therefore expect that
the fraction of extreme oxygen-rich stars are a non-negligible percentage of the
whole sample of O-rich stars of intermediate mass. 

We see in Fig.~\ref{fall} that a few thermal pulses take place each 2-4 kyr 
during the phase when the stars experience intense mass loss. 
Dust formation is temporarily halted during 
thermal pulses as the star readjusts on a more compact configuration, so as to counterbalance 
the effects of the CNO activity turning off, and the larger photospheric temperatures 
inhibit the formation of silicates in the wind. The stars interpreted as currently
experiencing strong mass loss can be evolving through any of the inter-pulse phases
of the super-wind evolution. Therefore, while the phase during which the stars
are expected to produce dust with large rates lasts a few thousand years, we expect that the dust 
responsible for the IR excess observed nowadays was not released earlier 
than 2-3 kyr ago, that is when the last TP took place. This 
understanding is consistent with the results by \citet{devries14}, who found that
the phase of intense dust production in these stars did not start earlier than
200-1000 yr ago, and was preceded by a phase during which dust production was
negligible.

Presently, we may consider the possibility that part of the most extreme OH$/$IR stars in the Milky
Way are single objects currently evolving through the final AGB stages, and the
remaining fraction belong to binary systems. Further observations of a wider
sample of this class of objects are required before it can be assessed whether
binarity is in fact a common rule for these objects and, more generally, to establish the fraction
of OH$/$IR stars that are part of binary systems.

\begin{table*}
\caption{Dust yields (solar masses) from stars of a different mass by various
research groups.}
\label{tabyield}      
\centering
\addtolength{\leftskip}{-2cm}
\addtolength{\leftskip}{-2cm}
\begin{tabular}{c c c c c}    
\hline      
${\rm M}/{\rm M}_{\odot}$ & This work & Ventura et al. (2018) & Nanni et al. (2013) & Ferrarotti \& Gail (2006) \\
\hline 
4 & $1.2\times 10^{-2}$ & $2.9\times 10^{-3}$ & $4\times 10^{-4}$    & $1.2\times 10^{-2}$   \\ 
5 & $2.1\times 10^{-2}$ & $4.5\times 10^{-3}$ & $9.3\times 10^{-3}$  & $1.5\times 10^{-2}$   \\
6 & $2.1\times 10^{-2}$ & $6.2\times 10^{-3}$ & $1.4\times 10^{-2}$ &  -                   \\
7 & $2.5\times 10^{-2}$ & $8.4\times 10^{-3}$ &  -                  & $1.8\times 10^{-2}$   \\
\hline
\end{tabular}
\end{table*}

\section{A revision of the dust yields from intermediate mass AGBs?}
\label{yields}
From the results shown in the bottom left panel of Fig.~\ref{fall}, it is
clear that the predictions regarding dust production between the VW93
and the Blo95 models are significantly different.
The Blo95 dust production rates are higher than VW93 for most of the AGB 
lifetime. However, this has very little effect on the overall dust yield of intermediate mass stars because most of the dust is released during the phases when the mass-loss rate is largest: during these phases, the VW93 $\dot{\rm M}_{\rm dust}$ values are
a factor $\sim 3$ higher than the Blo95 ones.

For the three models discussed in the previous section, we find that
the dust yields are $0.012~{\rm M}_{\odot}$ for ${\rm M}=4~{\rm M}_{\odot}$
and $0.021~{\rm M}_{\odot}$ for ${\rm M}=5,6~{\rm M}_{\odot}$. The dust yield
for the $7~{\rm M}_{\odot}$ model star (not shown for clarity concerns in
Fig.~\ref{fall}) is $0.025~{\rm M}_{\odot}$.
The dust released by the stars is mostly composed of silicates, which
account for $85\%$, with smaller contributions from alumina dust ($10\%$) 
and solid iron ($5\%$). These percentages are similar to those found
by \citet{ventura18}; on the other hand, the overall dust budget expected using VW95 mass loss is $\sim 3$ times larger than that of  \citet{ventura18} in which the Blo95 prescription was adopted.

In Table~\ref{tabyield} the dust yields from intermediate mass stars of
solar metallicity found in the present work are compared with those published
in \citet{ventura18}, and with results from \citet{nanni13} and 
\citet{fg06}\footnote{In the comparison with results from different studies,
it must be considered that the metallicity adopted here (Z=0.014) is smaller
than the value used in \citet{ventura18} (Z=0.018) and that used by
\citet{nanni13} and \citet{fg06} ($Z=0.02$)}. Our yields are
the largest among the solar metallicity dust yields published so far,
indicating that the contribution to the overall silicate dust production
by intermediate mass stars has been underestimated so far. We note that
the largest amount of carbon dust in \citet{ventura18}, released
by the $3~{\rm M}_{\odot}$ model star, is $\sim 0.015~{\rm M}_{\odot}$;
therefore, we find that at solar metallicities the quantity of
silicates produced by intermediate mass AGB stars is slightly higher than 
the amount of carbon that formed in the wind of low-mass stars.

The main results of the present
work concern the mass-loss rates suffered by intermediate mass
stars during the final AGB evolutionary stages, during and after the 
phases close to the luminosity peak experienced by these stars.
With the VW93 mass-loss prescription, this pertains to the AGB evolution from the
point when the superwind sets in.

Regardless of the prescription of mass-loss adopted, the effects on
the gas yields are negligible. The time between the
activation of HBB and the achievement of the luminosity peak is 
sufficiently long, such that the surface chemistry of the stars reached
the equilibrium distribution of the proton-capture nucleosynthesis
corresponding to the temperature attained at the bottom of the 
convective mantle. In this context, the chemical composition of
the gas ejected into the interstellar medium is independent on
the timescale with which the gas itself is released, which is determined by
 the mass-loss rate. We may safely conclude that the present
investigation has led us to pose some questions regarding the reliability of the
predicted silicate yields from intermediate mass
stars, but no conclusions regarding the gas yields from these objects
can be drawn based on the present analysis.

\section{Conclusions}
\label{conc}
We have studied a sample of Galactic AGB stars, whose SEDs exhibit
deep absorption features centred at $10~\mu$m and $18~\mu$m,
associated with silicate dust. The study of these stars, for which
the pulsation periods and some information on the surface chemical
composition are available, is based on stellar evolution and dust 
formation modelling, and supported by results from radiative 
transfer modelling. These ingredients have been used to characterise the 
individual objects in an attempt to determine the mass and formation 
epoch of the progenitors, and, on more general grounds, to understand
the silicate budget expected from AGB stars.

The sources investigated are interpreted as the progeny of intermediate
mass stars, currently evolving through advanced AGB phases, either
currently experiencing HBB or after HBB turned off by the gradual 
loss of the external envelope. The study of their SED indicates 
mass-loss rates in the $1-2\times 10^{-4}~{\rm M}_{\odot}/$yr
range, consistent with the radiation pressure wind description
in which photons experience multiple scattering processes.
These rates of mass loss of intermediate mass AGBs of solar 
metallicity are significantly higher (a factor $\sim 3$) than those 
published.
%so far by different research groups.
 
 Provided that this picture is correct, we deduce that mass loss
suffered by intermediate mass stars and the dust production mechanism 
remain highly efficient until the very final AGB phases, preceding
the contraction to the post-AGB phase.
These results indicate the need of an upwards revision of the
dust yields by intermediate mass stars, which are now found in
the $0.012-0.025~{\rm M}_{\odot}$ range, mostly under the form 
of silicates, with a smaller contribution from alumina dust 
($\sim 10\%$) and solid iron ($\sim 5\%$).

\label{end}

\begin{acknowledgements}
EM acknowledges support from the INAF research project 'LBT - Supporto Arizona Italia'. D.K. acknowledges  the  support  of  the  Australian  Research Council (ARC)  Discovery  Early  Career  Research  Award (DECRA) grant (DE190100813) and the Australian Research Council Centre of Excellence for All Sky Astrophysics in 3 Dimensions (ASTRO 3D), through project number CE170100013. MF acknowledges financial support from the ASI-INAF agreement n. 2022-14-HH.0. This research has made use of the GaiaPortal catalogues access tool, ASI?Space Science Data Center, Rome, Italy (http://gaiaportal.ssdc.asi.it).

\end{acknowledgements}


\end{document}